\documentclass{article}
\pdfoutput=1

\usepackage{arxiv}
\usepackage[utf8]{inputenc} 
\usepackage[T1]{fontenc} 
\usepackage{hyperref} 
\hypersetup{
	colorlinks = true,
	linkcolor = black,
	anchorcolor = black,
	citecolor = black,
	filecolor = black,
	urlcolor = blue
}
\usepackage{booktabs} 
\usepackage{amsfonts} 
\usepackage{nicefrac} 
\usepackage{microtype} 
\usepackage{amsmath,bm}
\usepackage{graphicx}
\usepackage[flushleft]{threeparttable}
\usepackage{multirow}
\usepackage{natbib}
\usepackage{makecell}
\usepackage{caption}
\usepackage{subcaption}   
\usepackage{nicematrix}
\usepackage{float}
\usepackage{rotating}

\newcommand{\figurehere}[1]{\begin{center}%
		=========================\\%
		Insert Figure #1 about here\\%
		=========================\\%
\end{center}}
\newcommand{\tablehere}[1]{\begin{center}%
		=========================\\%
		Insert Table #1 about here\\%
		=========================\\%
\end{center}}

\usepackage{array}
\newcommand{\PreserveBackslash}[1]{\let\temp=\\#1\let\\=\temp}
\newcolumntype{C}[1]{>{\PreserveBackslash\centering}p{#1}}
\newcolumntype{R}[1]{>{\PreserveBackslash\raggedleft}p{#1}}
\newcolumntype{L}[1]{>{\PreserveBackslash\raggedright}p{#1}}

\title{Decomposing Impact on Longitudinal Outcome of Time-varying Covariate into Baseline Effect and Temporal Effect}

\author{
  Jin Liu \thanks{CONTACT Jin Liu Email: Veronica.Liu0206@gmail.com}\\
Department of Biostatistics\\
Virginia Commonwealth University 
}

\begin{document}

\maketitle
\begin{abstract}
Longitudinal processes are often associated with each other over time; therefore, it is important to investigate the associations among developmental processes and understand their joint development. The latent growth curve model (LGCM) with a time-varying covariate (TVC) provides a method to estimate the TVC's effect on a longitudinal outcome while simultaneously modeling the outcome's change. However, it does not allow the TVC to predict variations in the random growth coefficients. We propose decomposing the TVC's effect into initial trait and temporal states using three methods to address this limitation. In each method, the baseline of the TVC is viewed as an initial trait, and the corresponding effects are obtained by regressing random intercepts and slopes on the baseline value. Temporal states are characterized as (1) interval-specific slopes, (2) interval-specific changes, or (3) changes from the baseline at each measurement occasion, depending on the method. We demonstrate our methods through simulations and real-world data analyses, assuming a linear-linear functional form for the longitudinal outcome. The results demonstrate that LGCMs with a decomposed TVC can provide unbiased and precise estimates with target confidence intervals. We also provide \textit{OpenMx} and \textit{Mplus 8} code for these methods with commonly used linear and nonlinear functions.
\end{abstract}


\section{Introduction}\label{Intro}
Longitudinal data analysis is a valuable tool for examining between-individual differences in within-individual changes across multiple disciplines, such as psychology, education, behavioral sciences, and biomedical sciences. In longitudinal studies, two or more sets of repeated measurements are often collected simultaneously. For instance, when assessing intelligence development, academic performance records across multiple subjects are often recorded over time for a comprehensive evaluation. Similarly, clinical trials repeatedly measure biomarkers and patient-reported outcomes (PROs) to assess treatment effects holistically. A compelling research topic involves the assessment of two or more longitudinal variables simultaneously, aiming to understand each process and the association of these processes.

Earlier studies have demonstrated multiple statistical models within the latent growth curve modeling (LGCM) framework to analyze joint developments. For example, \citet{McArdle1988Multi} proposed a parallel process and correlated growth model, also referred to as a multivariate growth model (MGM) in \citet{Grimm2007multi}, to explore two or more longitudinal variables simultaneously by estimating intercept-intercept and slope-slope covariances of two linear longitudinal processes. More recent studies have extended such MGMs with linear trajectories to investigate associations between multiple nonlinear longitudinal processes \citep{Blozis2004MGM, Blozis2008MGM, Liu2021PBLSGM} and the heterogeneity of these associations \citep{Liu2021PBLSGMM}. Assuming unidirectional rather than bidirectional relationships between-process random coefficients (also referred to as `growth factors' in the LGCM framework), researchers have also proposed analyzing joint development through longitudinal mediation models in the LGCM framework for linear \citep{Cheong2003mediate, Soest2011mediate, MacKinnon2008mediate} and nonlinear longitudinal processes \citep{Liu2022mediate}. Technical details and applications of these two types of statistical models for multiple longitudinal processes are well-documented in the publications above.

An additional model for examining joint development is the LGCM with a time-varying covariate (TVC), which estimates the effect of the TVC on a longitudinal outcome while simultaneously modeling change patterns in the longitudinal outcome. In contrast to the LGCM with a time-invariant covariate (TIC), where the covariate is assumed to remain constant over time and is used to explain the variance of random coefficients such as intercepts and slopes, the existing LGCM with a TVC enables more efficient use of data. However, it does not allow the covariates to account for variability in the random coefficients (e.g., random intercepts and slopes for a longitudinal model with a linear function). In this study, we propose decomposing the impact of a TVC into baseline and temporal effects to address this limitation. Specifically, we consider the baseline of the TVC as the initial trait and view (1) interval-specific slopes, (2) interval-specific changes, or (3) change-from-baseline values as temporal states. Each set of temporal states, combined with the initial trait, can capture the change patterns of a TVC. The regressions of the growth factors of the longitudinal outcome on the initial trait allow for the examination of baseline effects, while the regressions of the measurements of the longitudinal outcome on each set of temporal states enable the assessment of temporal effects.

In this section, we first discuss the inclusion of a covariate, which could be either a time-invariant covariate (TIC) or a time-varying covariate (TVC), when analyzing longitudinal data. We then describe the latent change score modeling (LCSM) framework \citep{Zhang2012LCSM, Grimm2013LCSM, Grimm2013LCM2} with the novel specification proposed by \citet{Liu2022LCSM}, which captures the change patterns of a longitudinal variable by estimating the baseline status (an initial trait) and the interval-specific slopes (a set of temporal states). Furthermore, we illustrate how to obtain the other two possible types of temporal states, the amount of change in each time interval and the change-from-baseline value at each measurement occasion, by modifying the specification in \citet{Liu2022LCSM} to include additional latent variables for the interval-specific changes or the change-from-baseline values.

\subsection*{Existing Longitudinal Model with Covariates}
The latent growth curve modeling (LGCM) and mixed-effects modeling frameworks are commonly employed tools for analyzing longitudinal data, focusing on individual changes over time and potential differences across these individual changes. The two modeling frameworks are mathematically and empirically equivalent in most cases \citep{Bauer2003equal, Curran2003equal}. In the LGCM framework, within-individual changes are analyzed through a set of growth factors (i.e., the random coefficients in the mixed-effects modeling framework) that together define the growth curve, such as an intercept and a slope for a linear change pattern. Between-individual differences are captured by the variances of these growth factors. LGCMs allow for the addition of time-invariant covariates (TICs) \citep{Joreskog1975MIMIC, McArdle1987MIMIC} to explain the variability of growth factors. TICs, like independent variables in a regression model, can be continuous or categorical, enabling the assessment of the conditional means and variance-covariance matrix of growth factors for the longitudinal outcome. TICs are individual-level variables that remain constant over time, such as biological sex, experimental condition, or individual-level assessments recorded only at baseline. The examination of how the between-individual differences in growth factors relate to these TICs provides insights into the possible reasons for between-individual differences in trajectories.

The LGCM with a TVC proposed by \citet{Grimm2007multi} can be fit within the mixed-effects modeling framework and the structural equation modeling (SEM) framework, with the models constructed in the SEM framework being able to provide more insights due to the flexibility of the framework. For example, the model in the SEM framework allows for varying effects of a TVC over time. It also enables one to estimate the covariances between the TVC and the growth factors of the longitudinal outcome \citep{Grimm2007multi}. However, the model also has limitations. First, the full model carries many parameters since there is no restricted structure on the TVC. Therefore, the mean vector, variance-covariance matrix, and TVC residuals must be estimated. The covariances among the TVC are impossible to estimate under some challenging conditions. One remedy for this is to assume the covariances to be zero, yet this assumption may not be valid if the TVC is expected to be stable to some extent \citep[Chapter~8]{Grimm2016growth}. Second, the model does not allow the TVC to predict variation in the growth factors of the longitudinal outcome. Third, the insight regarding the TVC effects from this existing model is limited to how the absolute values of the TVC affect the absolute values of the longitudinal outcome. However, researchers may also want to investigate how the change in a TVC affects the absolute value of the longitudinal outcome over time. \citet{Howard2015TVC} successfully addressed the first limitation by decomposing a TVC with the person-mean centering strategy. In particular, this strategy decomposes a TVC into the mean of a set of observed repeated measurements over time and the mean deviation for each individual. This work also demonstrated how to incorporate a decomposed TVC in the mixed-effects modeling framework.

The decomposition method proposed by \citet{Howard2015TVC} enables the evaluation of the effects of individual centers and mean deviations, providing meaningful interpretations when the trait of the TVC is assumed to be stable over time (at least within the study duration) and not influenced by any interventions. Conceptually, individual centers can be viewed as relatively stable traits under these assumptions, while the mean deviations represent situational states. Consequently, this decomposition allows for the assessment of both trait and temporal effects of the TVC. For instance, \citet{Howard2015TVC} applied this method to data from a 14-day daily diary study, successfully identifying the temporal effect of hours spent on schoolwork on the hours of sleep on the same day. Furthermore, this work demonstrated that such temporal effects differ between male and female university students. However, traits are not necessarily fixed and may change due to interventions, development, or learning experiences \citep{Steyer2015trait, Roberts2017trait}. To address this conceptual dilemma, \citet{Steyer2015trait} proposed time-specific common traits to analyze changes in traits. Building on the notion that traits change over time, we introduce three methods to decompose a TVC into an initial trait and a set of temporal states, addressing the limitations in previous research on TVCs \citep{Grimm2007multi, Howard2015TVC}, with the assumption that the TVC being examined is a `trait-like' variable but is sensitive to interventions, development, or learning experiences.

\subsection*{Modeling Framework Used for Decomposition of Time-Varying Covariate}
\citet{Liu2022LCSM} proposed a new specification for the LCSM framework. In addition to allowing researchers to build up an LCSM in the framework of individual measurement occasions, a feature of this new specification is that it allows for the estimation of the baseline value and individual interval-specific slopes. This characteristic provides a natural way to decompose a longitudinal variable into an initial trait and a collection of temporal states. Furthermore, this novel specification can be modified to derive other types of temporal states, such as individual interval-specific changes and individual change-from-baseline values, based on the estimated individual interval-specific slopes. Employing the LCSM with this novel specification and these possible modifications to describe a TVC reduces the number of parameters by fixing its structure. More importantly, it decomposes the initial trait and temporal states of the TVC. The estimated individual baseline value is then considered a predictor of growth factors of the longitudinal outcome, similar to a TIC. Meanwhile, the estimated individual interval-specific slopes, individual interval-specific changes, or individual change-from-baseline values can serve as possible predictors of the observed measurements of the longitudinal outcome. Although LCSMs with the novel specification can estimate change patterns for nonparametric and parametric nonlinear trajectories, we focus on the nonparametric curves in this section, as they are more relevant to the project. The novel specification for nonlinear parametric curves can also be employed to examine the change patterns of a TVC, which will be described in more detail in the Discussion section.

The nonparametric LCSM can also be viewed as the latent basis growth model (LBGM) fit within the LCSM framework \citep{Liu2022LCSM}. An LBGM consists of two growth factors: an intercept and a shape factor. There are multiple ways to scale the shape factor. Following \citet{Liu2022LCSM}, we consider the slope in the first time interval as the shape factor in this study; thus, the slope in each of the other time intervals can be viewed as the product of the shape factor and the corresponding relative rate. A path diagram of the model with six repeated measures is provided in Figure \ref{fig:path_slp}, where we use $x_{1}$-$x_{6}$ to represent repeated measures of the TVC in this project. As shown in Figure \ref{fig:path_slp}, the observed value at each measurement time $t_{j}$ is the sum of the corresponding latent true score (i.e., $x^{\ast}_{j}$) and a residual (i.e., $\epsilon_{xj}$). At baseline, the true score is the growth factor indicating the initial status (i.e., $\eta^{[x]}_{0}$). At each post-baseline, the true score at time $t_{j}$ is a linear combination of the score at the previous time point $t_{j-1}$ and the amount of true change from time $t_{j-1}$ to $t_{j}$, which is represented by the product of the time interval from $t_{j-1}$ to $t_{j}$ (i.e., $t_{j}-t_{j-1}$) and the slope (i.e., $dx_{j}$) in the interval. Furthermore, each interval-specific slope is the product of the shape factor (i.e., $\eta^{[x]}_{1}$) and the corresponding relative rate $\gamma_{j-1}$, as demonstrated in the figure. Note that the model specified in Figure \ref{fig:path_slp} allows for separately estimating the initial status and the interval-specific slopes. In Figure \ref{fig:path_slp}, the intervals from $t_{j-1}$ to $t_{j}$ are represented by diamond shapes, signifying that these time intervals are `definition variables' \citep{Mehta2000people, Mehta2005people, Sterba2014individually}. These are observed variables that adjust model coefficients to individual-specific values, allowing for model construction within the framework of individual measurement occasions.

\figurehere{1}

The parameters to be estimated in the model specified in Figure \ref{fig:path_slp} include the mean vector and variance-covariance matrix of the two growth factors, interval-specific relative rates, and residuals over time. An LBGM with the novel specification contains additional latent variables, such as true scores (i.e., $x^{\ast}_{j}$) at each measurement occasion and interval-specific slopes (i.e., $dx_{j}$), in addition to the two growth factors. However, these additional latent variables are not freely estimable; instead, they are derived from other parameters. Such non-estimable latent variables can be added into paths and then serve as predictors in a model within the SEM framework. We then view the true baseline as the initial trait and the interval-specific slopes as the set of temporal states for a TVC. We examine the baseline effect by regressing the growth factors of the longitudinal outcome on the estimated baseline value of the TVC and assess the temporal effect by regressing each measurement of a longitudinal outcome on the slope of the previous time interval of the TVC.

The model in Figure \ref{fig:path_slp} can be modified by adding additional latent variables to allow for different types of temporal states. In Figure \ref{fig:path_chg}, we add $\delta x_{j}$ to represent the amount of change in the time interval from $t_{j-1}$ to $t_{j}$. Similarly, in Figure \ref{fig:path_chgBL}, we add $\Delta x_{j}$ to represent the change-from-baseline at time $t_{j}$. Similar to the true scores $x^{\ast}_{j}$ and slopes $dx_{j}$, $\delta x_{j}$ and $\Delta x_{j}$ are derived from other parameters rather than being freely estimated. They can also be included in paths and serve as predictors in a model. So we have three possible methods to decompose a TVC. In all three methods, the estimated TVC baseline value is the initial trait, while the interval-specific slopes, interval-specific changes, or change-from-baseline values represent three possible types of temporal states. In this project, we demonstrate the proposed methods within the framework of individual measurement occasions, following multiple existing studies that illustrate this framework for LGCMs \citep{Sterba2014individually, Liu2019BLSGM} and LCSMs \citep{Grimm2018Individually, Liu2022LCSM} to avoid potential inadmissible solutions \citep{Blozis2008coding}.

\figurehere{2}

The remainder of this article is structured as follows. First, we outline the model specification and estimation for each of the three proposed decomposition methods. In the subsequent section, we detail the design of a Monte Carlo simulation to evaluate the performance of these methods. Specifically, we present performance metrics, including relative bias, empirical standard error (SE), relative root-mean-squared-error (RMSE), and empirical coverage probability (CP) for a nominal $95\%$ confidence interval of parameters of interest. In the Application section, we analyze a real-world dataset to illustrate the existing and proposed methods. Finally, we conclude with a Discussion section that addresses practical and methodological considerations, as well as potential avenues for future research.

\section*{Method}\label{Method}
\subsection*{Decomposition of Time-varying Covariate}
This section details the three methods to decompose a TVC into an initial trait and a set of temporal states. We begin with the novel specification for the latent basis growth model (LBGM) introduced by \citet{Liu2022LCSM}, which enables a decomposition of a TVC into the estimated baseline value (i.e., an initial trait) and interval-specific slopes (i.e., a collection of temporal states). In line with multiple earlier studies, such as \citet{McArdle2001LCM1} and \citet[Chapter~11]{Grimm2016growth}, \citet{Liu2022LCSM} views the LBGM with $J$ measurements as a linear piecewise function with $J-1$ segments. For the $i^{th}$ individual, the model can be specified as follows
\begin{align}
&x_{ij}=x^{\ast}_{ij}+\epsilon^{[x]}_{ij},\label{eq:LBGM1}\\
&x^{\ast}_{ij}=\begin{cases}
\eta^{[x]}_{0i}, & \text{if $j=1$}\\
x^{\ast}_{i(j-1)}+dx_{ij}\times(t_{ij}-t_{i(j-1)}), & \text{if $j=2, \dots, J$}
\end{cases},\label{eq:LBGM2_1}\\
&dx_{ij}=\eta^{[x]}_{1i}\times\gamma_{j-1}\qquad (j=2, \dots, J). \label{eq:LBGM3}
\end{align}

Equations \ref{eq:LBGM1}, \ref{eq:LBGM2_1}, and \ref{eq:LBGM3} together define a LBGM specified in Figure \ref{fig:path_slp}. In Equation \ref{eq:LBGM1}, $x_{ij}$, $x^{\ast}_{ij}$, and $\epsilon^{[x]}_{ij}$ represent the observed measurement, latent true score, and residual of individual $i$ at time $t_{j}$, respectively. Equation \ref{eq:LBGM2_1} demonstrates that the latent true scores have different expressions at baseline (i.e., $j=1$) and post-baseline (i.e., $j\ge2$). At baseline, the true score corresponds to the growth factor indicating the intercept (i.e., $\eta^{[x]}_{0i}$). At each post-baseline, the true score at time $t_{j}$ is a linear combination of the score at the previous time point $t_{j-1}$ and the amount of true change from time $t_{j-1}$ to $t_{j}$, which is the product of the time interval (i.e., $t_{j}-t_{(j-1)}$) and the slope (i.e., $dx_{ij}$) within that interval. In Equation \ref{eq:LBGM3}, the slope in each interval is further expressed by the product of the slope of the first interval (i.e., the shape factor $\eta^{[x]}_{1i}$) and the corresponding relative rate (i.e., $\gamma_{j-1}$).

The model specified in Equations \ref{eq:LBGM1}-\ref{eq:LBGM3} can be modified to accommodate additional latent variables that indicate interval-specific changes, as demonstrated in Equations \ref{eq:LBGM2_21} and \ref{eq:LBGM2_22}:
\begin{align}
&x^{\ast}_{ij}=\begin{cases}
\eta^{[x]}_{0i}, & \text{if $j=1$}\\
x^{\ast}_{i(j-1)}+\delta x_{ij}, & \text{if $j=2, \dots, J$}
\end{cases},\label{eq:LBGM2_21}\\
&\delta x_{ij}=dx_{ij}\times(t_{ij}-t_{i(j-1)})\qquad (j=2, \dots, J), \label{eq:LBGM2_22}
\end{align}
where $\delta x_{ij}$ indicates the amount of change from $t_{j-1}$ to $t_{j}$ for the $i^{th}$ individual. Thus, Equations \ref{eq:LBGM1}, \ref{eq:LBGM2_21}, \ref{eq:LBGM2_22}, and \ref{eq:LBGM3} together define an LBGM specified in Figure \ref{fig:path_chg}. The only difference between this modified specification and the original specification proposed in \citet{Liu2022LCSM} is the explicit inclusion of interval-specific changes in the model specified by Equations \ref{eq:LBGM1}, \ref{eq:LBGM2_21}, \ref{eq:LBGM2_22}, and \ref{eq:LBGM3}, allowing these interval-specific changes to serve as predictors within the SEM framework.

Similarly, we modify the model specified in Equations \ref{eq:LBGM1}-\ref{eq:LBGM3} to include additional latent variables that indicate the change-from-baseline values, as shown in Equations \ref{eq:LBGM2_31} and \ref{eq:LBGM2_32}:
\begin{align}
&x^{\ast}_{ij}=\begin{cases}
\eta^{[x]}_{0i}, & \text{if $j=1$}\\
\eta^{[x]}_{0i}+\Delta x_{ij}, & \text{if $j=2, \dots, J$}
\end{cases},\label{eq:LBGM2_31}\\
&\Delta x_{ij}=\Delta x_{i(j-1)}+dx_{ij}\times(t_{ij}-t_{i(j-1)})\qquad (j=2, \dots, J). \label{eq:LBGM2_32}
\end{align}
Here, $\Delta x_{ij}$ indicates the change-from-baseline at time $t_{j}$ for individual $i$, which is a linear combination of the change-from-baseline at the earlier time point $t_{j-1}$ and the amount of change from $t_{j-1}$ to $t_{j}$. Equations \ref{eq:LBGM1}, \ref{eq:LBGM2_31}, \ref{eq:LBGM2_32}, and \ref{eq:LBGM3} together define an LBGM specified in Figure \ref{fig:path_chgBL}. The true score at each time point $t_{j}$ can be expressed as the sum of the baseline value and the corresponding change-from-baseline. By explicitly including them in the model specification, these change-from-baseline values are allowed to be predictors.

The novel specification proposed in \citet{Liu2022LCSM} and the two possible modifications can be expressed in the same matrix form, which is only related to the freely estimable parameters in the model,
\begin{equation}\nonumber
\boldsymbol{x}_{i}=\boldsymbol{\Lambda}^{[x]}_{i}\times\boldsymbol{\eta}^{[x]}_{i}+\boldsymbol{\epsilon}^{[x]}_{i}
\end{equation}
in which $\boldsymbol{x}_{i}$ is a $J\times1$ vector of the repeated measurements of the TVC of individual $i$ (where $J$ is the number of measurement occasions), and $\boldsymbol{\eta}^{[x]}_{i}$ is a $2\times1$ vector of the growth factors of the TVC, representing the initial status and the slope of the first time interval for individual $i$, respectively. Furthermore, $\boldsymbol{\Lambda}^{[x]}_{i}$ is a $J\times2$ matrix of the corresponding factor loadings,
\begin{equation}\nonumber
\boldsymbol{\Lambda}^{[x]}_{i}=\begin{pmatrix}
1 & 0 \\
1 & \gamma_{1}\times(t_{i2}-t_{i1}) \\
1 & \sum_{j=2}^{3}\gamma_{j-1}\times(t_{ij}-t_{i(j-1)}) \\
\dots & \dots \\
1 & \sum_{j=2}^{J}\gamma_{j-1}\times(t_{ij}-t_{i(j-1)}) \\
\end{pmatrix}.
\end{equation}
The elements of the first column are all $1$, as they are the factor loadings of the TVC intercept. The $j^{th}$ element of the second column is the cumulative value of the relative rate up to time $t_{j}$, so its product with $\eta^{[x]}_{1i}$ represents the change-from-baseline value at time $t_{j}$. Additionally, $\boldsymbol{\epsilon}^{[x]}_{i}$ is a $J\times1$ vector of residuals for individual $i$. The growth factors $\boldsymbol{\eta}^{[x]}_{i}$ can be further expressed as:
\begin{equation}\nonumber \boldsymbol{\eta}^{[x]}_{i}=\boldsymbol{\mu}^{[x]}_{\boldsymbol{\eta}}+\boldsymbol{\zeta}^{[x]}_{i}, \end{equation}
where $\boldsymbol{\mu}^{[x]}$ is the mean vector of the TVC growth factors, and $\boldsymbol{\zeta}^{[x]}_{i}$ is the vector of deviations of the $i^{th}$ individual from the corresponding growth factor means. More technical details can be found in \citet{Liu2022LCSM}.

\subsection*{Model Specification of Latent Growth Curve Model with Decomposed Time-varying Covariate}
This section presents the model specification of the proposed LGCMs with a decomposed TVC, in which the growth factors of the longitudinal outcome are regressed on the initial trait, while each post-baseline observed measurement of the longitudinal outcome is regressed on the corresponding value of each of the three types of temporal states. In this section, we do not pre-specify any functional form for the longitudinal outcome; instead, we provide a general model specification that can be adapted to any growth curve function.

For the $i^{th}$ individual, the proposed LGCMs with a TIC and a TVC that is decomposed into its baseline value and interval-specific slopes can be expressed as
\begin{equation}
\begin{pmatrix}\boldsymbol{x}_{i} \\ \boldsymbol{y}_{i}
\end{pmatrix}=\begin{pmatrix}
\boldsymbol{\Lambda}^{[x]}_{i} & \boldsymbol{0} \\
\boldsymbol{0} & \boldsymbol{\Lambda}^{[y]}_{i}
\end{pmatrix}\times\begin{pmatrix} \boldsymbol{\eta}^{[x]}_{i} \\ \boldsymbol{\eta}^{[y]}_{i}
\end{pmatrix}+\kappa_{1}\times\begin{pmatrix} \boldsymbol{0} \\ \boldsymbol{dx_{i}}
\end{pmatrix}+\begin{pmatrix}
\boldsymbol{\epsilon}^{[x]}_{i} \\ \boldsymbol{\epsilon}^{[y]}_{i}
\end{pmatrix}, \label{eq:LGCM1_1}   
\end{equation}
where $\boldsymbol{y}_{i}$ is a $J\times1$ vector of the repeated measures of the $i^{th}$ individual (with $J$ being the number of measurement occasions), $\boldsymbol{\eta}^{[y]}_{i}$ is a $K\times1$ vector of growth factors (where $K$ is the number of growth factors of the longitudinal outcome), $\boldsymbol{\Lambda}^{[y]}_{i}$ is a $J\times K$ matrix of the corresponding factor loadings (with the subscript $i$ indicating that the model is constructed with individual measurement occasions), and $\boldsymbol{\epsilon}^{[y]}_{i}$ is a $J\times1$ vector of residuals for individual $i$. Additionally, $\boldsymbol{dx_{i}}$ is a $J\times1$ vector of interval-specific slopes of the TVC, which can be further expressed as $\boldsymbol{dx_{i}}=\begin{pmatrix}0 & dx_{i2} & dx_{i3} & \dots & dx_{iJ}\end{pmatrix}$, where the first element is $0$ and $dx_{ij}$ is the slope in the $(j-1)^{th}$ time interval of the $i^{th}$ individual. Therefore, in Equation \ref{eq:LGCM1_1}, $\kappa_{1}$ is the temporal effect of the TVC, which indicates how the value of $\boldsymbol{y}_{i}$ at $t_{j}$ is affected by the slope in the previous time interval (i.e., from $t_{j-1}$ to $t_{j}$). In the above equation, $\boldsymbol{0}$ is a $J\times1$ vector, and other notations have the same definitions as in previous equations.

We then further regress the growth factors on the TIC and the latent true score of the initial status (i.e., the growth factor that indicates the intercept or the true baseline value) of the TVC
\begin{equation}
\boldsymbol{\eta}^{[y]}_{i}=\boldsymbol{\alpha}^{[y]}+\begin{pmatrix}\boldsymbol{\beta_{\text{TIC}}} & \boldsymbol{\beta_{\text{TVC}}}\end{pmatrix}\times\begin{pmatrix}X_{i} \\ \eta^{[x]}_{0i}\end{pmatrix} +\boldsymbol{\zeta}^{[y]}_{i}, \label{eq:LGCM2}
\end{equation}
in which $\boldsymbol{\alpha}^{[y]}$ is a $K\times1$ vector of growth factor intercepts, $\boldsymbol{\beta_{\text{TIC}}}$ is a $K\times1$ vector of regression coefficients from the TIC to the growth factors, and $\boldsymbol{\beta_{\text{TVC}}}$ is a $K\times1$ vector of regression coefficients from the estimated initial status of the TVC to the growth factors. Additionally, $X_{i}$ is the TIC value and $\eta^{[x]}_{0i}$ is the TVC estimated initial status of the $i^{th}$ individual, and $\boldsymbol{\zeta}^{[y]}_{i}$ is a $K\times1$ vector of deviations of individual $i$ from the conditional means of growth factors.

Similarly, the proposed LGCMs with a TIC and a TVC with the other two types of decomposition for individual $i$ can be expressed in Equations
\begin{equation}\label{eq:LGCM1_2}
\begin{pmatrix}\boldsymbol{x}_{i} \\ \boldsymbol{y}_{i}
\end{pmatrix}=\begin{pmatrix}
\boldsymbol{\Lambda}^{[x]}_{i} & \boldsymbol{0} \\
\boldsymbol{0} & \boldsymbol{\Lambda}^{[y]}_{i}
\end{pmatrix}\times\begin{pmatrix} \boldsymbol{\eta}^{[x]}_{i} \\ \boldsymbol{\eta}^{[y]}_{i}
\end{pmatrix}+\kappa_{2}\times\begin{pmatrix} \boldsymbol{0} \\ \boldsymbol{\delta x_{i}}
\end{pmatrix}+\begin{pmatrix}
\boldsymbol{\epsilon}^{[x]}_{i} \\ \boldsymbol{\epsilon}^{[y]}_{i}
\end{pmatrix},     
\end{equation}
and 
\begin{equation}\label{eq:LGCM1_3}
\begin{pmatrix}\boldsymbol{x}_{i} \\ \boldsymbol{y}_{i}
\end{pmatrix}=\begin{pmatrix}
\boldsymbol{\Lambda}^{[x]}_{i} & \boldsymbol{0} \\
\boldsymbol{0} & \boldsymbol{\Lambda}^{[y]}_{i}
\end{pmatrix}\times\begin{pmatrix} \boldsymbol{\eta}^{[x]}_{i} \\ \boldsymbol{\eta}^{[y]}_{i}
\end{pmatrix}+\kappa_{3}\times\begin{pmatrix} \boldsymbol{0} \\ \boldsymbol{\Delta x_{i}}
\end{pmatrix}+\begin{pmatrix}
\boldsymbol{\epsilon}^{[x]}_{i} \\ \boldsymbol{\epsilon}^{[y]}_{i}
\end{pmatrix},
\end{equation}
respectively, where $\boldsymbol{\delta x_{i}}$ is a $J\times1$ vector of interval-specific changes of TVC, which can be further expressed as $\boldsymbol{\delta x_{i}}=\begin{pmatrix}0 & \delta x_{i2} & \delta x_{i3} & \dots & \delta x_{iJ}\end{pmatrix}$, while $\boldsymbol{\Delta x_{i}}$ is a $J\times1$ vector of change-from-baseline values of TVC, which can be further expressed as $\boldsymbol{\Delta x_{i}}=\begin{pmatrix}0 & \Delta x_{i2} & \Delta x_{i3} & \dots & \Delta x_{iJ}\end{pmatrix}$. Similar to $\boldsymbol{dx_{i}}$, the first element of the two vectors is $0$, and $\delta x_{ij}$ and $\Delta x_{ij}$ are the amount of change in the $(j-1)^{th}$ time interval and the change-from-baseline at $j^{th}$ time point of individual $i$, respectively. Therefore, both $\kappa_{2}$ and $\kappa_{3}$ can be interpreted as a temporal effect of the TVC, which demonstrates how the observed measurement of $\boldsymbol{y}_{i}$ at $t_{j}$ is affected by the amount of change in the previous time interval and change-from-baseline at $t_{j}$, respectively. We can further regress $\boldsymbol{\eta}^{[y]}_{i}$ in Equations \ref{eq:LGCM1_2} and \ref{eq:LGCM1_3} on the TIC and the latent true score of the TVC baseline value as we did for $\boldsymbol{\eta}^{[y]}_{i}$ in Equation \ref{eq:LGCM1_1}. The regressions of $\boldsymbol{\eta}^{[y]}_{i}$ have the same expression as Equation \ref{eq:LGCM2}.

In practice, the longitudinal outcome $\boldsymbol{y}_{i}$ could take either a linear or nonlinear function, depending on the trajectory shape demonstrated by raw data and the research questions of interest. In Table \ref{tbl:traj_summary}, we list functional forms, growth factors and corresponding interpretations, and factor loadings for the linear growth curve and four commonly used nonlinear trajectories.

\tablehere{1}

\subsection*{Model Estimation}
This section describes how to estimate LGCMs with a decomposed TVC. In order to simplify the estimation, we make the following assumptions. First, the TIC and the growth factors of TVC are normally distributed\footnote{Note that the normality assumption of TIC is not necessary when fitting the proposed models; here, we have this assumption to allow for the estimation of the covariance between the TIC and the estimated TVC initial value.}; that is, $X_{i}\sim N(\mu_{x}, \phi_{x})$ and $\boldsymbol{\zeta}^{[x]}_{i}\sim \text{MVN}(\boldsymbol{0}, \boldsymbol{\Phi}^{[x]}_{\boldsymbol{\eta}})$, where $\mu_{x}$ and $\phi_{x}$ are the mean and variance of the TIC, while $\boldsymbol{\Phi}^{[x]}_{\boldsymbol{\eta}}$ is a $2\times2$ variance-covariance matrix of the TVC growth factors. The second assumption is that the growth factors of the longitudinal outcome are normally distributed conditional on the TIC and the true score of the TVC initial status, that is, $\boldsymbol{\zeta}^{[y]}_{i}\sim \text{MVN}(\boldsymbol{0}, \boldsymbol{\Psi}^{[y]}_{\boldsymbol{\eta}})$, in which $\boldsymbol{\Psi}^{[y]}_{\boldsymbol{\eta}}$ is a $K\times K$ variance-covariance matrix of the growth factors of the longitudinal outcome. We also assume that the residuals of the TVC and the longitudinal outcome are identical and independent normal distributions and that the residual covariances are homogeneous over time, that is,
\begin{equation}\nonumber
\begin{pmatrix}
\boldsymbol{\epsilon}^{[x]}_{i} \\ \boldsymbol{\epsilon}^{[y]}_{i}\end{pmatrix}\sim \text{MVN}\bigg(\begin{pmatrix}
\boldsymbol{0} \\ \boldsymbol{0}\end{pmatrix}, \begin{pmatrix} \theta^{[x]}_{\epsilon}\boldsymbol{I} &
\theta^{[xy]}_{\epsilon}\boldsymbol{I} \\ & \theta^{[y]}_{\epsilon}\boldsymbol{I} \end{pmatrix}\bigg),
\end{equation}
in which $\boldsymbol{I}$ is a $J\times J$ identity matrix. The expected mean vector and variance-covariance structure of the TVC and the longitudinal outcome for the proposed LGCMs with a decomposed TVC are provided in Appendix \ref{supp1}.

The parameters in the proposed LGCMs with a decomposed TVC include the mean ($\mu_{x}$) and variance ($\phi_{x}$) of the TIC, the mean vector ($\boldsymbol{\mu}^{[x]}_{\boldsymbol{\eta}}$) and variance-covariance matrix ($\boldsymbol{\Phi}^{[x]}_{\boldsymbol{\eta}}$) of the TVC growth factors, the intercepts ($\boldsymbol{\alpha}^{[y]}$) and unexplained variance-covariance ($\boldsymbol{\Psi}^{[y]}_{\boldsymbol{\eta}}$) of the growth factors of the longitudinal outcome, the effects of the TIC ($\boldsymbol{\beta}_{\text{TIC}}$) on the growth factors of the longitudinal outcome, and the baseline effects ($\boldsymbol{\beta}_{\text{TVC}}$), the temporal effect of the TVC ($\kappa_{1}$, $\kappa_{2}$, or $\kappa_{3}$ for the three types of decomposition, respectively), and residual variances ($\theta^{[x]}$ and $\theta^{[y]}$) and covariance ($\theta^{[xy]}$). Additionally, the relationship between the TIC and TVC is captured by the correlation ($\rho_{\text{BL}}$) between the TIC and the initial trait of the TVC (i.e., the true score of the TVC initial status). We define $\boldsymbol{\Theta}$ as
\begin{equation}\nonumber
\begin{aligned}
\boldsymbol{\Theta} = \{ &\mu_{x}, \phi_{x}, \boldsymbol{\mu}^{[x]}_{\boldsymbol{\eta}}, \boldsymbol{\Phi}^{[x]}_{\boldsymbol{\eta}}, \gamma_{2}, \dots, \gamma_{J-1}, \\
&\boldsymbol{\alpha}^{[y]}, \boldsymbol{\Psi}^{[y]}_{\boldsymbol{\eta}}, \boldsymbol{\beta}_{\text{TIC}}, \boldsymbol{\beta}_{\text{TVC}}, \kappa_{1/2/3}, \rho_{\text{BL}}, \theta^{[x]}, \theta^{[y]}, \theta^{[xy]}\}
\end{aligned}
\end{equation}
to list the parameters specified in the model. Additionally, as demonstrated in Table \ref{tbl:traj_summary}, there may exist additional growth coefficients in LGCMs, such as the coefficient $b$ in the negative exponential trajectory, $c$ in the Jenss-Bayley growth curve, and $\gamma$ in the bilinear spline function. These additional growth coefficients also need to be estimated in the proposed models.

In this study, the full information maximum likelihood (FIML) technique (i.e., the MLE in the statistical literature) was used to estimate the proposed LGCMs with a decomposed TVC to account for the heterogeneity of individual contributions to the likelihood. We used the R package \textit{OpenMx} with the optimizer CSOLNP \citep{OpenMx2016package, Pritikin2015OpenMx, Hunter2018OpenMx, User2020OpenMx} to build the proposed models. We provide \textit{OpenMx} code in the online appendix (\url{https://github.com/xxxx}) (code will be uploaded upon acceptance) to demonstrate how to employ the proposed novel specification. The proposed LGCMs with a decomposed TVC can also be fit using other SEM software such as \textit{Mplus 8}. We also provide the corresponding code on the GitHub website for researchers interested in using it.

\section*{Model Evaluation}\label{Evaluate}
In this project, we conducted a Monte Carlo simulation study to evaluate the proposed methods for TVC decomposition. For three reasons, we assumed the longitudinal outcome takes the bilinear spline functional form with an unknown fixed knot, as described in Table \ref{tbl:traj_summary}. First, the inclusion of a TVC when fitting an LGCM breaks the variance of the longitudinal outcome into three parts: the variance explained by growth factors, the variance explained by temporal states of a TVC, and an unexplained part (i.e., the residual). Therefore, we hypothesized that adding a TVC into a model may affect the estimation of the growth factors of the longitudinal outcome to some extent. This linear-linear piecewise function allows us to investigate how the inclusion of temporal states of the TVC affects slope estimation in the earlier and later stages of the longitudinal outcome. Second, the trajectory of the longitudinal outcome could take any linear or nonlinear functional form in the LGCM framework, yet the simulation result observed from the model with one functional form is easy to extend to other functions. Third, earlier studies such as \citet{Liu2019BLSGM} have documented the results of extensive simulation studies for this functional form, which further reduces the size of simulation conditions in this project as we can only focus on how the size of trait and temporal effects affects model performance. We examined the three decomposition methods through four performance measures listed in Table \ref{tbl:metric}, where we also include definitions and estimates of these four measures.

\tablehere{2}

When designing the simulation, we decided on the number of repetitions $S=1,000$ using an empirical method suggested by \citet{Morris2019simulation}. Specifically, we performed a pilot study and observed that the standard errors of all coefficients were less than $0.15$, except for the initial status of the TVC and the longitudinal outcome, which requires at least $900$ repetitions to keep the Monte Carlo standard error of the bias less than $0.15$\footnote{Bias is the most important performance metric in a simulation study, and the equation of its Monte Carlo standard error is $\text{Monte Carlo SE(Bias)}=\sqrt{Var(\hat{\theta})/S}$ \citep{Morris2019simulation}.}. We then proceeded with the simulation with $1,000$ replications for more conservative consideration.

\subsection*{Design of Simulation Study}\label{Simu:design}
All the conditions considered for all three decomposition methods in the simulation design are provided in Table \ref{tbl:simu_design}. In general, we manipulated the conditions presumed to affect TVC effects while keeping the rest fixed to limit the size of the simulation conditions in this project. The number of measurement occasions is important when examining longitudinal processes, as a longitudinal model presumably performs better under conditions with more repeated records. We are interested in investigating whether the number of repeated measures affects the TVC effects through this simulation study. To this end, two levels of repeated measurements, six and ten waves, were selected in the design. The condition with six measures was chosen for model identification purposes \citep{Bollen2005LGC, Liu2019BLSGM}, while the other condition with ten measurements was selected to examine whether a longer study duration would improve model performance. Additionally, we set individual measurement occasions by allowing a `medium' time window $(-0.25, +0.25)$ around each wave following \citet{Coulombe2015ignoring}. 

\tablehere{3}

In the simulation design, we considered standardized TIC and fixed the distribution of the growth factors of the longitudinal outcome and the TVC, as examining how the trajectory shapes affect model performance is beyond this project's scope. We considered a midway knot for the trajectories of the longitudinal outcome. The correlation between the TIC and the initial trait of the TVC was set as $0.3$. Three levels of baseline effects were included in the simulation design to allow the TIC and the initial trait of the TVC to account for $13\%$ or $26\%$ variability of the growth factors of the longitudinal outcome. In the first scenario, all coefficients from the TVC initial trait were set as $0$, and the TIC was designated to explain $13\%$ variability of the growth factors of the longitudinal outcome. The TIC and the initial state of the TVC jointly accounted for $13\%$ or $26\%$ variability in the second and third scenarios, respectively. Furthermore, we considered four levels of temporal effect for each of the three methods. With these conditions for the regression coefficients, our objective was to determine whether the proposed methods could detect trait and temporal effects. Additionally, we set the residual covariance between the longitudinal outcome and the TVC at a moderate level (i.e., the correlation was set as $0.3$) and considered two levels of sample size ($n=200$ or $500$).

\subsection*{Data Generation and Simulation Step}
The simulation study for each condition of each method listed in Table \ref{tbl:simu_design} was conducted through the following steps:

\begin{enumerate}
\item {Generate the TIC, the growth factors of the TVC, and the growth factors of the longitudinal outcome for an LGCM with a decomposed TVC using the R package \textit{MASS} \citep{Venables2002Statistics},}
\item {Generate a time structure with $J$ waves $t_{j}$ ($J=6$ or $10$) and allow for disturbances around each wave $t_{ij}\sim U(t_{j}-\Delta, t_{j}+\Delta)$ ($\Delta=0.25$) to establish individual measurement occasions, }
\item {Compute factor loadings for the longitudinal outcome and the TVC, which are functions of the individual measurement occasions and additional growth coefficient(s) (i.e., the knot for the longitudinal outcome and the relative rates of the TVC),}
\item {Calculate a set of temporal states of the TVC, which includes interval-specific slopes (from the slope in the first time interval and relative rates), interval-specific changes (from the interval-specific slopes and individual time intervals), or change-from-baseline (from the accumulative interval-specific changes) for the three methods, respectively,}
\item {Determine the true values of the repeated measures for the TVC and the longitudinal outcome: the former is based on its growth factors and factor loadings, while the latter relies on its growth factors, factor loadings, and a set of temporal states of the TVC; then add the residual matrix to the longitudinal outcome and the TVC,}
\item {Fit the LGCM with each decomposition method, estimate the parameters, and construct the corresponding $95\%$ Wald confidence intervals,}
\item {Repeat the above steps until achieving $1,000$ convergent solutions.}
\end{enumerate}

\section*{Results}\label{Result}
In this section, we summarize the simulation results. First, we examined the convergence\footnote{Convergence in this project is defined as achieving the \textit{OpenMx} status code $0$, which suggests successful optimization.} rate of each LGCM with a decomposed TVC. All three proposed methods with a linear-linear piecewise longitudinal outcome exhibited excellent convergence, as they reported a $100\%$ convergence rate for all conditions listed in Table \ref{tbl:simu_design}.

Next, we evaluated the performance metrics of each parameter for the three decomposition methods under all conditions, including relative bias, empirical SE, relative RMSE, and empirical coverage of the nominal $95\%$ confidence interval. Given the size of the simulation conditions and the number of parameters, we first examined the summary statistics of each of the four performance metrics for each parameter of each decomposition method. Specifically, we calculated each performance metric across $1000$ replications under each condition and summarized the results across all conditions into the corresponding median and range. Our simulation results indicate that the first two decomposition methods generated unbiased and accurate point estimates with target probabilities. The summary statistics of the four performance measures for these two methods are provided in Tables S1 and S2 in the online supplementary document, respectively.

However, the third decomposition method demonstrated some bias in the estimates of mean values of the slopes of the longitudinal outcome (especially the slope in the later stage), the baseline effects on these two slopes, and the temporal effect (i.e., the relative bias could reach $10\%$, see Table S3), leading to unsatisfactory performance in relative RMSE and coverage probability. We then plotted the relative biases of these parameters stratified by the number of repeated measures, sample size, size of baseline and temporal effects, and residual variance of the longitudinal outcome in Figures S1-S5. From these figures, we observed that such biased estimates were produced under conditions with shorter study durations (i.e., $J=6$). Furthermore, the estimate of the temporal effect was more biased under conditions with larger baseline or smaller temporal effects; the latter might be attributed to the small population value (i.e., $\kappa_{3}=0.2$).

One possible explanation for the poor performance of the third decomposition method could be the large size of its temporal states (the change-from-baseline) compared to the temporal states in the other two methods (i.e., the interval-specific slopes or interval-specific changes). As mentioned earlier, we regress the longitudinal outcome on these state characteristics; thus, the larger temporal states might lead to a reduction in the size of the temporal effect while simultaneously inflating the estimates of coefficients related to the growth factors of the longitudinal outcome, especially under conditions with a shorter study duration. This phenomenon could also explain the increased bias in the estimation of the mean value of the second slope of the longitudinal outcome, as the temporal states are even larger during the later stage.

Therefore, based on our simulation study, the first two decomposition methods performed satisfactorily under all examined conditions, while the third method only performed well under certain mild conditions. This suggests that the third method requires meticulous data preprocessing and model interpretation in practice, which we will demonstrate in the Application section.

\section*{Application}\label{Application}
This section demonstrates how to employ the proposed TVC decomposition methods to analyze bivariate longitudinal variables, which are viewed as a TVC and a longitudinal outcome, respectively, with baseline characteristics (i.e., TICs). The application section has two goals. The first goal is to provide feasible recommendations for employing the proposed TVC decomposition methods to answer specific research questions. The second goal is to show how the inclusion of a TVC when modeling a LGCM affects the estimation of growth factors in real-world practice. To achieve this aim, we built LGCMs with the three proposed methods and the existing LGCM with a TVC. We randomly selected $400$ students from the Early Childhood Longitudinal Study, Kindergarten Cohort: 2010-2011 (ECLS-K: 2011) with non-missing records of repeated reading and mathematics assessments with baseline teacher-reported inhibitory control for this application\footnote{ECLS-K: 2011 contains $n= 18174$ participants. There are $n=3144$ students after removing rows with missing values ((i.e., records with any of $NaN/-9/-8/-7/-1$).}. 

ECLS-K: 2011 is a longitudinal study that starts from the 2010-2011 school year and collects records from children enrolled in approximately $900$ kindergarten programs across the United States. In the survey, students' reading and mathematics abilities were assessed in nine waves: each semester in kindergarten, first and second grade, followed by only the spring semester in third, fourth, and fifth grade. As pointed out by \citet{Le2011ECLS}, only about one-third of students were evaluated in the 2011 and 2012 fall semesters. In this analysis, we used the age-in-month for each wave so that each student had individual measurement occasions. 

We fit the LGCMs with a decomposed TVC and the existing LGCM with a TVC to analyze how students' baseline teacher-reported inhibitory control and reading ability affect their development of mathematics ability with the assumption that the mathematics trajectories take the bilinear spline function with an unknown fixed knot. We standardized the TIC, baseline teacher-reported inhibitory control. For the TVC, reading achievement scores over time, we first calculated the mean and variance of the baseline reading ability and then standardized the ability at each wave using the baseline mean and variance. All four models converged, and the estimated likelihood, Akaike information criterion (AIC), Bayesian information criterion (BIC), residual variance, and the number of parameters of each model are provided in Table \ref{tbl:info}. In addition to these four models with a TVC, we constructed two reference models, the LGCM and LGCM with a TIC only, and the corresponding model summary is also included in Table \ref{tbl:info}. From the table, we noticed that adding a TVC into an LGCM increased the estimated likelihood, and then AIC and BIC, but did not affect the residual variance of mathematics growth curves meaningfully. Among the four models with a TVC, the LGCM with decomposed TVC into an initial trait and change-from-baseline values had the smallest estimated likelihood, AIC, and BIC, suggesting that this model fit the raw data best from a statistical perspective.

\tablehere{4}

To examine how the inclusion of a TVC affects the estimation of the growth factors of mathematics achievement, we plotted a model-implied curve on the smooth line of the development of mathematics ability for each of the six models in Figure \ref{fig:curves}. The figure shows that the estimated trajectory from the models without a TVC fits the smooth line. The development of mathematics ability can be viewed in two stages: the growth in the early stage is relatively rapid and then slows down substantially around $110$ months old. The growth factors of mathematics development from the models incorporating a TVC were somewhat underestimated, which aligns with our expectations. When a TVC is included in the LGCM as a predictor of the observed measurement for the longitudinal outcome, it leads to smaller estimates of the growth factors. This effect is more evident in the later stage, as the increased reading ability over time exerts a greater influence on the estimation. On the contrary, the impact on the estimation of growth factors for interval-specific slopes or changes is relatively minor, so the growth factors are able to provide a more accurate representation of the growth patterns. The estimates of these two models are presented in Tables \ref{tbl:est_TVCslp} and \ref{tbl:est_TVCchg}, respectively.

\figurehere{3}

\tablehere{5}

\tablehere{6}

Table \ref{tbl:est_TVCslp} shows that, for the development of mathematics ability, the estimated knot was around $ 9$ years old and that the pre-and post-knot growth rates were $1.62$ and $0.64$, respectively. As stated earlier, we standardized reading scores at each wave with the baseline mean and variance. The estimated mean of the baseline score was $0.06$, which is within our expectation since the score at baseline has been centered to $0$ when performing the standardization. The estimated slope mean during the first interval was $0.17$, indicating that the standardized reading ability increased $0.17$ per month during Grade K. The other interval-specific slopes can be calculated from the slope in the first interval and the corresponding relative rates. For example, the slope during Grade 1 was $0.14$ (i.e., $0.17\times0.81$). More detailed interpretations of the growth coefficients for the TVC and the longitudinal outcome can be found in earlier research works, such as \citet{Liu2019BLSGM} and \citet{Liu2022LCSM}.

After decomposing the reading ability development, we can also estimate the covariance between the TIC and the TVC initial trait, which was $0.25$ in this application. This suggests that teacher-reported inhibitory control and baseline reading ability were positively associated. In addition, one standardized unit increase in the baseline reading score resulted in a $6.07$ and $0.09$ increase in the initial status and the first slope of the mathematics ability development. The estimated temporal effect was $27.37$, indicating that, for example, one unit increase in the slope of standardized reading ability development during Grade K led to a $27.37$ increase in mathematics final scores at the end of Grade K.

The estimates from the model with the decomposed TVC into the initial trait and interval-specific changes are similar to those shown in Table \ref{tbl:est_TVCchg}, except for the temporal effect. The temporal effect of this model was $4.37$, suggesting that, for example, one unit increase in the change of standardized reading ability development during Grade K led to a $4.37$ increase in final mathematics scores at the end of Grade K.

\section*{Discussion}\label{Discussion}
This article proposes three methods for decomposing a TVC to evaluate the baseline and temporal effects separately. Specifically, we treat the estimated baseline value as the initial trait and either the interval-specific slopes, interval-specific changes, or change-from-baseline values as a set of temporal states. We assessed the three decomposition methods through extensive simulation studies by fitting LGCMs with a decomposed TVC, assuming that the longitudinal outcome follows a bilinear spline function with an unknown fixed knot. Based on our simulation studies, the LGCMs with the first two decomposition methods are capable of estimating the parameters of interest unbiasedly and accurately with generating target coverage probabilities. However, the estimates from the model with the third method might be biased under challenging conditions, such as those with a shorter duration. The patterns we observed in the simulation study were also supported by real-world data analysis: the underestimation of the growth factors of the longitudinal outcome with the first two decomposition methods was negligible, while this underestimation of the third method was more evident. 

\subsection*{Practical Considerations}\label{D:practical}
In this section, we provide a set of recommendations for empirical researchers based on the simulation study and real-world data analysis. First, we proposed three TVC decomposition methods in this article; therefore, along with the existing methods, there are five approaches for incorporating a TVC when analyzing longitudinal data, with four developed in the LGCM framework. Our aim is not to demonstrate that a TVC should be added to a longitudinal model in a decomposed way or that one decomposed method is universally preferable. The selection of methods should be based on the nature of the TVC, research objectives, and specific research questions. 

For instance, if the research interest is to estimate the trait and state effects separately, longitudinal models with a decomposed TVC are suitable candidates. If the TVC of interest is a `state-like' variable (for example, a person's mood), which is temporary and sensitive to external situations but typically not responsive to interventions, development, or learning experiences, a decomposed TVC into the person-mean and mean deviation may be a good choice. Conversely, if the TVC under examination is a `trait-like' variable (e.g., a person's cognitive skills, such as reading ability), which is more stable across situations but sensitive to interventions, development, or learning experiences, a decomposed TVC into the initial trait and temporal states should be considered. Moreover, the decomposition methods that produce the baseline effect and interval-specific slopes or changes as the temporal effect are appropriate for obtaining unbiased estimates and growth factors that reflect developmental processes. Additionally, the estimated effect size of the temporal effect and the corresponding interpretation vary across methods, as demonstrated in the Application section. This aspect also needs to be considered when selecting a method. For example, if the research question is to understand how the growth rate of reading ability affects mathematics achievement, the decomposition method with the interval-specific slopes is an ideal candidate model. Similarly, the method with the interval-specific changes helps examine how the amount of change in reading ability influences mathematics ability.

Second, the results of real-world data analysis show that a good index value (e.g., AIC or BIC) for a model does not necessarily guarantee unbiased estimates or meaningful interpretations of the estimated parameters that accurately reflect developmental processes. For example, the LGCM with the third decomposition method has the smallest AIC and BIC values but underestimates slopes, especially the post-knot slope of mathematics development. Consequently, we recommend fitting LGCMs without a TVC as reference models, as illustrated in the Application section. As evidenced by the simulation study in prior research, such as \citet{Liu2019BLSGM}, LGCMs without a TVC can successfully capture the underlying change patterns of a longitudinal outcome. Thus, reference models without a TVC help understand any discrepancies between the LGCM with a TVC and the change patterns evident in the raw data.

Third, researchers often standardize a covariate when including it in a model to make the estimated effects comparable across covariates. However, when standardizing a TVC, it is essential to exercise caution to preserve the change patterns and the variance-covariance structure of the TVC. One approach is to standardize the TVC at each wave using a constant mean and variance, such as scaling the TVC across time with the mean and variance of its baseline values. After fitting the model, this standardization strategy enables the transformation of the estimated results into effects in the original scale for interpretation purposes.

Lastly, the simulation study indicates that the estimates of coefficients related to slopes and the temporal effect from the LGCM with a TVC decomposed into an initial trait and change-from-baseline values might exhibit some bias, especially under challenging conditions. As a result, interpreting the mean values of slopes, baseline effects on slopes, and the temporal effect should be approached with caution. The simulation study suggests that the coefficients related to slopes could be overestimated, while the temporal effect could be underestimated.

\subsection*{Methodological Considerations and Future Directions}\label{D:method}
There are several directions to consider for future studies. First, including the temporal effects of a TVC as a predictor of the corresponding observed measurements of the longitudinal outcome can result in estimated growth factor means that are lower than the actual change patterns, with influences that vary depending on the size of the temporal states. This observation helps explain the poor performance of the third decomposition method, where change-from-baseline values, which are larger than the interval-specific slopes or changes and accumulate over time, serve as the temporal states. In the proposed methods, we assume that the temporal effect is homogeneous over time to simplify model specification and enable the use of these methods within the mixed-effects modeling framework. Relaxing this assumption may improve the performance of the third method, as the squeezing impacts on the growth factors of the temporal states could be alleviated if the temporal effect is allowed to be adjusted with the magnitude of the corresponding temporal states. Examining methods with a relaxed assumption is beyond the scope of the present project but could be a future direction.

Second, the present study assumes that the TVC takes a nonparametric functional form, of which only the interval-specific slopes are of interest. In addition to the new specification for the nonparametric function, \citet{Liu2022LCSM} included this novel specification for multiple parametric nonlinear trajectories, such as quadratic and negative exponential functions. Apart from the interval-specific slopes, these parametric functions also allow for the examination of growth coefficients, such as growth acceleration or growth capacity. Therefore, these nonlinear parametric functional forms can also be used to describe the change patterns of a TVC if the evaluation of these coefficients of the TVC is also of research interest. The proposed methods can be extended accordingly.

Third, we conducted the simulation study for the proposed methods under the assumption that the longitudinal outcome follows a linear-linear functional form. This was done to demonstrate how temporal states affect the estimation of growth factors for the longitudinal outcome in both earlier and later stages separately. Although a LGCM with a decomposed TVC where the longitudinal outcome takes other functions is likely to exhibit similar patterns to those observed in this project, additional simulations can be performed to answer specific research questions, such as how temporal states affect the estimation of growth acceleration.

Fourth, we illustrated the proposed methods using the same time structure for the TVC and the longitudinal outcome. However, it is possible to extend these methods for a bivariate longitudinal process in which one variable has fewer measurements. Lastly, the proposed methods can also be extended to analyze a bivariate longitudinal process with dropout under the assumption of missing at random, thanks to the FIML technique used in model estimation.

\subsection*{Concluding Remarks}\label{D:conclude}
In summary, this article proposes three methods to decompose a TVC into an initial trait and a set of temporal states. Specifically, we consider the baseline value as the initial trait and the interval-specific slopes, changes, or change-from-baseline values as possible sets of temporal states. Using the proposed methods, we can evaluate the baseline effect and temporal effect of a TVC separately. As discussed earlier, these methods can be further extended in practice and further examined methodologically.


\newpage

\bibliographystyle{apalike}
\bibliography{Extension9}

\appendix
\renewcommand{\thesection}{Appendix \Alph{section}}
\renewcommand{\thesubsection}{A.\arabic{subsection}}
\section{Mean vector and Variance-covariance Matrix of the Longitudinal Variables}\label{supp1}
\renewcommand{\theequation}{A.\arabic{equation}}
\setcounter{equation}{0}

For the LGCM with a TVC that is decomposed to the baseline value and interval-specific slopes, the expected mean vector and variance-covariance structure of the TVC ($\boldsymbol{x}_{i}$) and the longitudinal outcome ($\boldsymbol{y}_{i}$) for the $i^{th}$ individual can be expressed as
\begin{equation}\nonumber
\boldsymbol{\mu}_{i}=\begin{pmatrix}
\boldsymbol{\mu}^{[x]}_{i} \\ \boldsymbol{\mu}^{[y]}_{i} 
\end{pmatrix}=\begin{pmatrix}
\boldsymbol{\Lambda}^{[x]}_{i} & \boldsymbol{0} \\ \boldsymbol{0} & \boldsymbol{\Lambda}^{[y]}_{i}
\end{pmatrix}\times\begin{pmatrix}
\boldsymbol{\mu}^{[x]}_{\boldsymbol{\eta}} \\ \boldsymbol{\mu}^{[y]}_{\boldsymbol{\eta}}
\end{pmatrix}+\kappa_{1}\times\begin{pmatrix} \boldsymbol{0} \\ \boldsymbol{\mu}_{dx} \end{pmatrix},
\end{equation}
and
\begin{equation}\nonumber
\boldsymbol{\Sigma}_{i}=\begin{pmatrix}
\boldsymbol{\Sigma}^{[x]}_{i} & \boldsymbol{\Sigma}^{[xy]}_{i} \\ & \boldsymbol{\Sigma}^{[y]}_{i}
\end{pmatrix}=\begin{pmatrix}
\boldsymbol{\Lambda}^{[x]}_{i} & \boldsymbol{0} \\ \boldsymbol{0} & \boldsymbol{\Lambda}^{[y]}_{i}
\end{pmatrix}\times \begin{pmatrix}
\boldsymbol{\Phi}^{[x]}_{\boldsymbol{\eta}} & \boldsymbol{0} \\
\boldsymbol{0} & \boldsymbol{\text{Var(y)}}
\end{pmatrix} \times\begin{pmatrix}
\boldsymbol{\Lambda}^{[x]}_{i} & \boldsymbol{0} \\ \boldsymbol{0} & \boldsymbol{\Lambda}^{[y]}_{i}
\end{pmatrix}^{T}+\begin{pmatrix}\boldsymbol{0} & \boldsymbol{0} \\ \boldsymbol{0} & \kappa^{2}_{1}\boldsymbol{\phi}_{dx}
\end{pmatrix}+\begin{pmatrix} \theta^{[x]}_{\epsilon}\boldsymbol{I} &
\theta^{[xy]}_{\epsilon}\boldsymbol{I} \\ & \theta^{[y]}_{\epsilon}\boldsymbol{I} \end{pmatrix},
\end{equation}
respectively, where $\boldsymbol{\mu}_{dx}$ and $\boldsymbol{\phi}_{dx}$ are the means and variances of the interval-specific slopes, which are not freely estimable parameters in the proposed model. In practice, such parameters can be created by the function \textit{mxAlgebra()} in \textit{OpenMx} and the estimates are stored in the corresponding objective while the standard errors are usually evaluated by the function \textit{mxSE()}. In addition, $\boldsymbol{\mu}^{[y]}_{\boldsymbol{\eta}}$ and $\boldsymbol{\text{Var(y)}}$ are the conditional mean vector and variance-covariance matrix of the growth factors of the longitudinal outcome on the TIC and the true score of the TVC initial status, which can be further expressed as
\begin{equation}\nonumber
\boldsymbol{\mu}^{[y]}_{\boldsymbol{\eta}}=\boldsymbol{\alpha}^{[y]}+\begin{pmatrix}\boldsymbol{\beta_{\text{TIC}}} & \boldsymbol{\beta_{\text{TVC}}}\end{pmatrix}\times\begin{pmatrix}\mu_{x} \\ \mu^{[x]}_{\eta_{0}}\end{pmatrix},
\end{equation}
and
\begin{equation}\nonumber
\boldsymbol{\text{Var(y)}}=\boldsymbol{\Psi}^{[y]}_{\boldsymbol{\eta}}+\begin{pmatrix}\boldsymbol{\beta_{\text{TIC}}} & \boldsymbol{\beta_{\text{TVC}}}\end{pmatrix}\times\begin{pmatrix} \phi_{x} & \rho_{\text{BL}}\sqrt{\phi_{x}\phi^{[x]}_{00}} \\
\rho_{\text{BL}}\sqrt{\phi_{x}\phi^{[x]}_{00}} & \phi^{[x]}_{00}
\end{pmatrix}\times\begin{pmatrix}\boldsymbol{\beta_{\text{TIC}}} & \boldsymbol{\beta_{\text{TVC}}}\end{pmatrix}^{T},
\end{equation}
respectively, where $\mu^{[x]}_{\eta_{0}}$ and $\phi^{[x]}_{00}$ are the mean and variance of the TVC initial status, and $\rho_{\text{BL}}$ is the correlation between the TIC and the true value of the TVC initial status.
\newpage

For the LGCM with a TVC that is decomposed into the baseline value and interval-specific changes, the expected mean vector and variance-covariance structure of the TVC ($\boldsymbol{x}_{i}$) and the longitudinal outcome ($\boldsymbol{y}_{i}$) for the $i^{th}$ individual can be expressed as
\begin{equation}\nonumber
\boldsymbol{\mu}_{i}=\begin{pmatrix}
\boldsymbol{\mu}^{[x]}_{i} \\ \boldsymbol{\mu}^{[y]}_{i} 
\end{pmatrix}=\begin{pmatrix}
\boldsymbol{\Lambda}^{[x]}_{i} & \boldsymbol{0} \\ \boldsymbol{0} & \boldsymbol{\Lambda}^{[y]}_{i}
\end{pmatrix}\times\begin{pmatrix}
\boldsymbol{\mu}^{[x]}_{\boldsymbol{\eta}} \\ \boldsymbol{\mu}^{[y]}_{\boldsymbol{\eta}}
\end{pmatrix}+\kappa_{2}\times\begin{pmatrix} \boldsymbol{0} \\ \boldsymbol{\mu}_{\delta x} \end{pmatrix},
\end{equation}
and
\begin{equation}\nonumber
\boldsymbol{\Sigma}_{i}=\begin{pmatrix}
\boldsymbol{\Sigma}^{[x]}_{i} & \boldsymbol{\Sigma}^{[xy]}_{i} \\ & \boldsymbol{\Sigma}^{[y]}_{i}
\end{pmatrix}=\begin{pmatrix}
\boldsymbol{\Lambda}^{[x]}_{i} & \boldsymbol{0} \\ \boldsymbol{0} & \boldsymbol{\Lambda}^{[y]}_{i}
\end{pmatrix}\times \begin{pmatrix}
\boldsymbol{\Phi}^{[x]}_{\boldsymbol{\eta}} & \boldsymbol{0} \\
\boldsymbol{0} & \boldsymbol{\text{Var(y)}}
\end{pmatrix} \times\begin{pmatrix}
\boldsymbol{\Lambda}^{[x]}_{i} & \boldsymbol{0} \\ \boldsymbol{0} & \boldsymbol{\Lambda}^{[y]}_{i}
\end{pmatrix}^{T}+\begin{pmatrix}\boldsymbol{0} & \boldsymbol{0} \\ \boldsymbol{0} & \kappa^{2}_{2}\boldsymbol{\phi}_{\delta x}
\end{pmatrix}+\begin{pmatrix} \theta^{[x]}_{\epsilon}\boldsymbol{I} &
\theta^{[xy]}_{\epsilon}\boldsymbol{I} \\ & \theta^{[y]}_{\epsilon}\boldsymbol{I} \end{pmatrix},
\end{equation}
respectively, where $\boldsymbol{\mu}_{\delta x}$ and $\boldsymbol{\phi}_{\delta x}$ are the means and variances of the interval-specific changes, which are not freely estimable parameters and can be derived using the function \textit{mxAlgebra()}. $\boldsymbol{\mu}^{[y]}_{\boldsymbol{\eta}}$ and $\boldsymbol{\text{Var(y)}}$ have the same definitions as the previous equations. 

For the LGCM with a TVC that is decomposed into the baseline value and change-from-baseline values, the expected mean vector and variance-covariance structure of the TVC ($\boldsymbol{x}_{i}$) and the longitudinal outcome ($\boldsymbol{y}_{i}$) for the $i^{th}$ individual can be expressed as
\begin{equation}\nonumber
\boldsymbol{\mu}_{i}=\begin{pmatrix}
\boldsymbol{\mu}^{[x]}_{i} \\ \boldsymbol{\mu}^{[y]}_{i} 
\end{pmatrix}=\begin{pmatrix}
\boldsymbol{\Lambda}^{[x]}_{i} & \boldsymbol{0} \\ \boldsymbol{0} & \boldsymbol{\Lambda}^{[y]}_{i}
\end{pmatrix}\times\begin{pmatrix}
\boldsymbol{\mu}^{[x]}_{\boldsymbol{\eta}} \\ \boldsymbol{\mu}^{[y]}_{\boldsymbol{\eta}}
\end{pmatrix}+\kappa_{3}\times\begin{pmatrix} \boldsymbol{0} \\ \boldsymbol{\mu}_{\Delta x} \end{pmatrix},
\end{equation}
and
\begin{equation}\nonumber
\boldsymbol{\Sigma}_{i}=\begin{pmatrix}
\boldsymbol{\Sigma}^{[x]}_{i} & \boldsymbol{\Sigma}^{[xy]}_{i} \\ & \boldsymbol{\Sigma}^{[y]}_{i}
\end{pmatrix}=\begin{pmatrix}
\boldsymbol{\Lambda}^{[x]}_{i} & \boldsymbol{0} \\ \boldsymbol{0} & \boldsymbol{\Lambda}^{[y]}_{i}
\end{pmatrix}\times \begin{pmatrix}
\boldsymbol{\Phi}^{[x]}_{\boldsymbol{\eta}} & \boldsymbol{0} \\
\boldsymbol{0} & \boldsymbol{\text{Var(y)}}
\end{pmatrix} \times\begin{pmatrix}
\boldsymbol{\Lambda}^{[x]}_{i} & \boldsymbol{0} \\ \boldsymbol{0} & \boldsymbol{\Lambda}^{[y]}_{i}
\end{pmatrix}^{T}+\begin{pmatrix}\boldsymbol{0} & \boldsymbol{0} \\ \boldsymbol{0} & \kappa^{2}_{3}\boldsymbol{\phi}_{\Delta x}
\end{pmatrix}+\begin{pmatrix} \theta^{[x]}_{\epsilon}\boldsymbol{I} &
\theta^{[xy]}_{\epsilon}\boldsymbol{I} \\ & \theta^{[y]}_{\epsilon}\boldsymbol{I} \end{pmatrix},
\end{equation}
respectively, where $\boldsymbol{\mu}_{\Delta x}$ and $\boldsymbol{\phi}_{\Delta x}$ are the means and variances of the interval-specific changes, which are not freely estimable parameters and can be derived using the function \textit{mxAlgebra()}. $\boldsymbol{\mu}^{[y]}_{\boldsymbol{\eta}}$ and $\boldsymbol{\text{Var(y)}}$ have the same definitions as the previous equations. 

\newpage


\renewcommand\thefigure{\arabic{figure}}
\setcounter{figure}{0}

\begin{figure*}
\centering
  \includegraphics[width=1\linewidth]{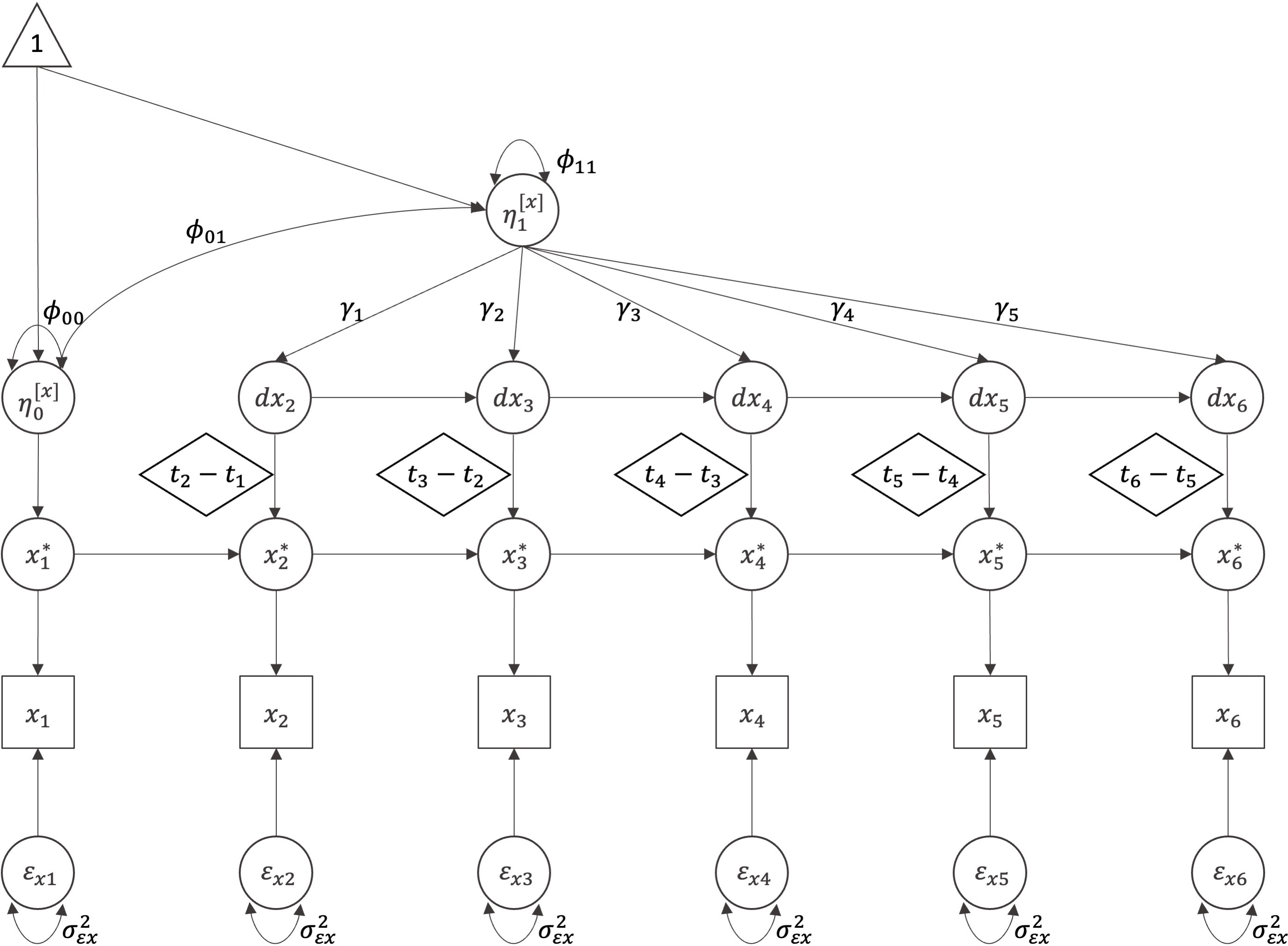}
\caption{Path Diagram of Latent Basis Growth Model with Novel Specification\\
Note: boxes=manifested variables, circles=latent variables, single arrow=regression paths;
doubled arrow=(co)variances; triangle=constant; diamonds=definition variables.\\
In the model, $\gamma_{1}$ is set as $1$ for model identification considerations.}
\label{fig:path_slp}
\end{figure*}

\begin{figure*}
\centering
\begin{subfigure}{.5\textwidth}
  \centering
  \includegraphics[width=1\linewidth]{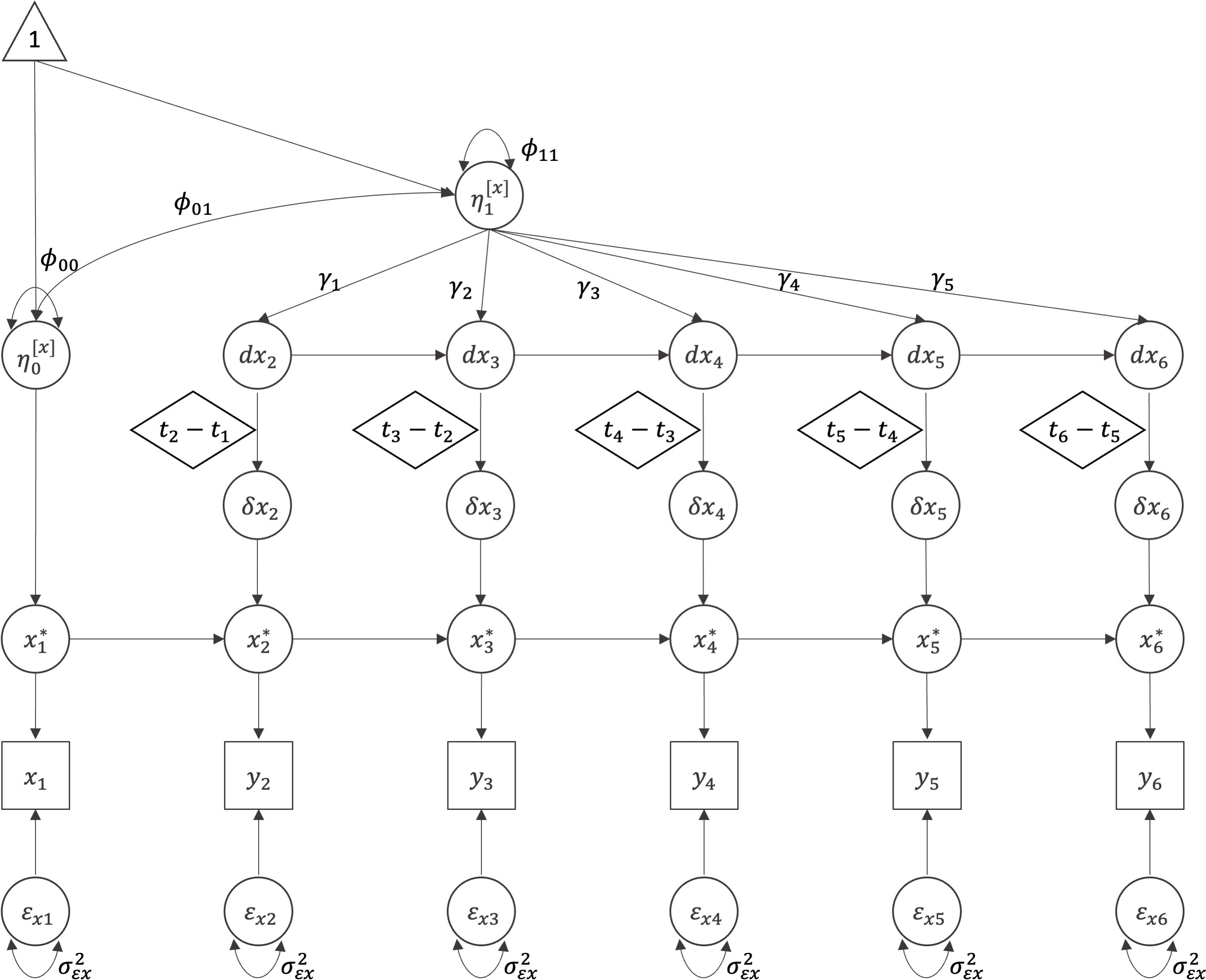}
  \caption{Estimating Change in Each Interval}
  \label{fig:path_chg}
\end{subfigure}%
\begin{subfigure}{.5\textwidth}
  \centering
  \includegraphics[width=1\linewidth]{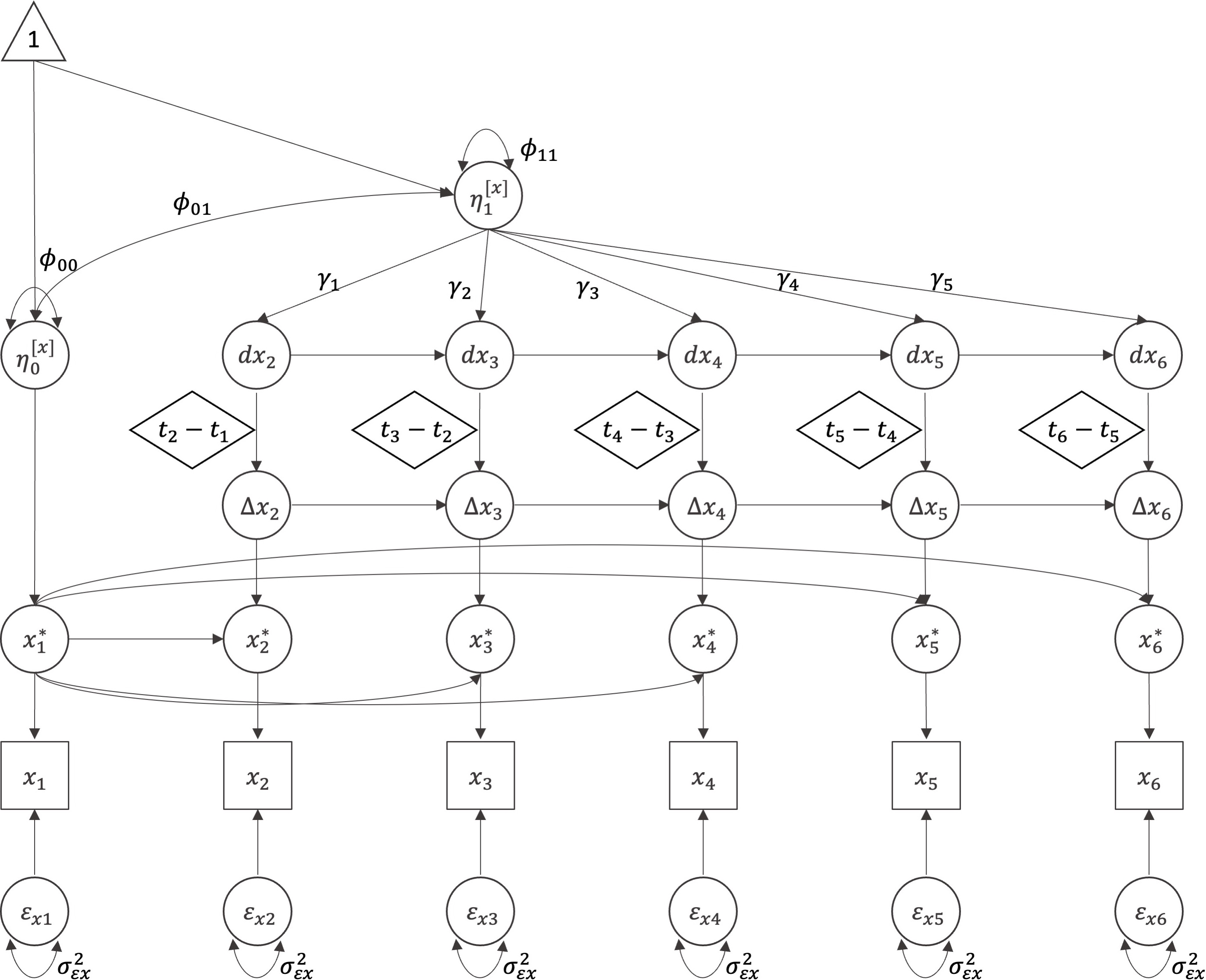}
  \caption{Estimating Change from Baseline}
  \label{fig:path_chgBL}
\end{subfigure}
\caption{Modified Path Diagram of Latent Basis Growth Model with Novel Specification\\
Note: boxes=manifested variables, circles=latent variables, single arrow=regression paths;
doubled arrow=(co)variances; triangle=constant; diamonds=definition variables.\\
In the model, $\gamma_{1}$ is set as $1$ for model identification considerations.}
\label{fig:path_ext}
\end{figure*}

\begin{figure*}
\centering
  \includegraphics[width=1\linewidth]{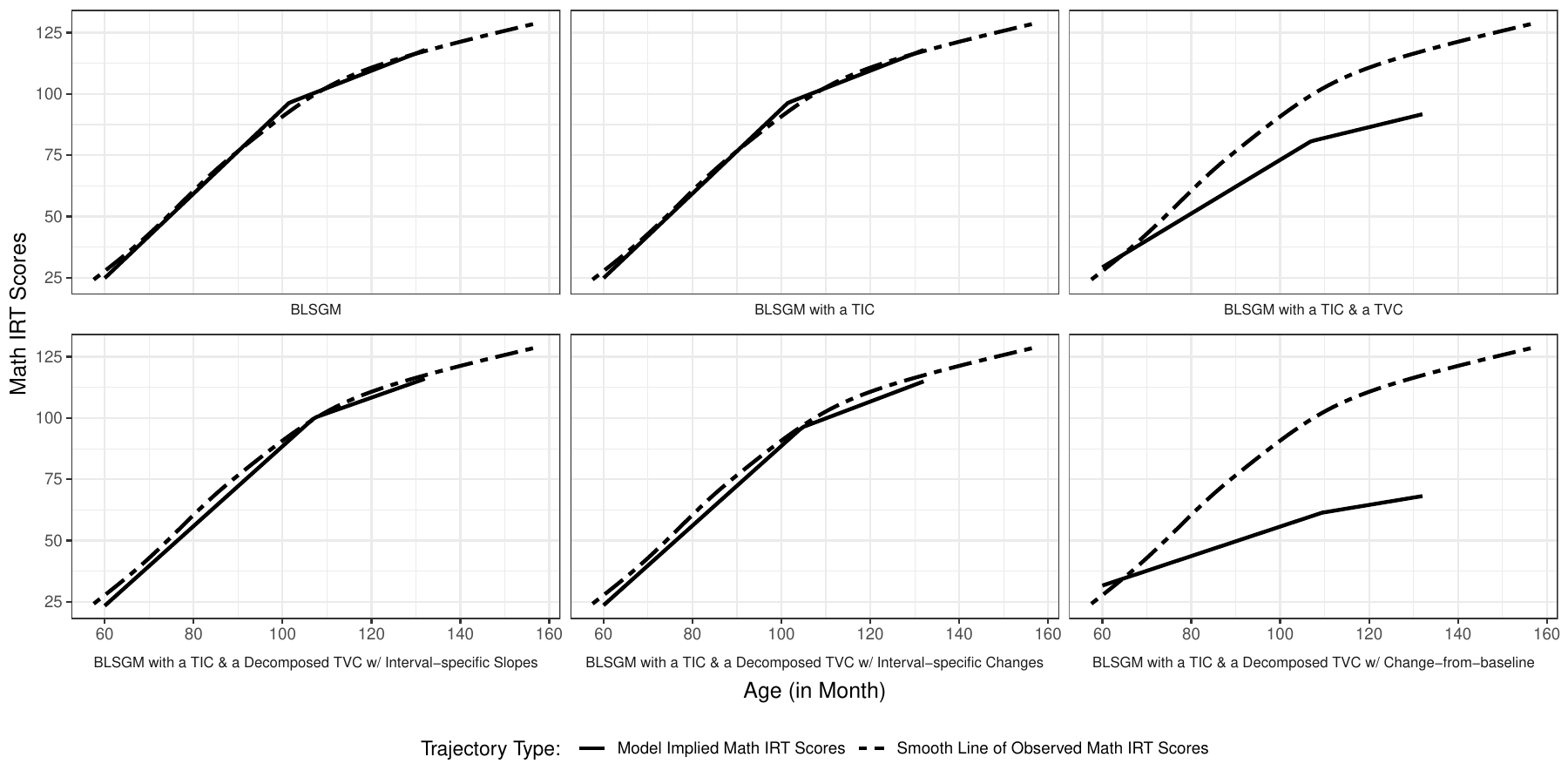}
\caption{Model Implied Trajectory and Smooth Line of Mathematics Performance\\
Note: BLSGM stands for bilinear spline growth model with an unknown fixed knot.}
\label{fig:curves}
\end{figure*}

\clearpage


\renewcommand\thetable{\arabic{table}}
\setcounter{table}{0}

\begin{sidewaystable*}
\centering
\footnotesize
\begin{threeparttable}
\setlength{\tabcolsep}{5pt}
\renewcommand{\arraystretch}{0.75}
\caption{Summary of Linear and Commonly Used Nonlinear Functional Forms}
\begin{tabular}{p{6cm}p{5cm}p{10cm}}
\hline
\hline
\textbf{Growth Factors}\tnote{a} & \textbf{Factor Loadings}\tnote{b} & \textbf{Interpretation of Growth Factors and Additional Coef.} \\
\hline
\multicolumn{3}{l}{\textbf{Linear Function}: $y_{ij}=\eta^{[y]}_{0i}+\eta^{[y]}_{1i}\times{t_{ij}}+\epsilon^{[y]}_{ij}$}\\
\hline
\multirow{2}{*}{$\boldsymbol{\eta}^{[y]}_{i}=\begin{pmatrix}\eta^{[y]}_{0i} & \eta^{[y]}_{1i}\end{pmatrix}$} & 
\multirow{2}{*}{$\boldsymbol{\Lambda}^{[y]}_{i}=\begin{pmatrix}1 & t_{ij} \end{pmatrix}$} & 
$\eta^{[y]}_{0i}$: the initial status \\
& & $\eta^{[y]}_{1i}$: the linear component of change \\
\hline
\hline
\multicolumn{3}{l}{\textbf{Quadratic Function}: $y_{ij}=\eta^{[y]}_{0i}+\eta^{[y]}_{1i}\times{t_{ij}}+\eta^{[y]}_{2i}\times{t^{2}_{ij}}+\epsilon^{[y]}_{ij}$}\\
\hline
\multirow{3}{*}{$\boldsymbol{\eta}^{[y]}_{i}=\begin{pmatrix}\eta^{[y]}_{0i} & \eta^{[y]}_{1i} & \eta^{[y]}_{2i}\end{pmatrix}$} & 
\multirow{3}{*}{$\boldsymbol{\Lambda}^{[y]}_{i}=\begin{pmatrix}1 & t_{ij} & t^{2}_{ij} \end{pmatrix}$} & 
$\eta^{[y]}_{0i}$: the initial status \\
& & $\eta^{[y]}_{1i}$: the linear component of change \\
& & $\eta^{[y]}_{2i}$: the quadratic component of change \\
\hline
\hline
\multicolumn{3}{l}{\textbf{Negative Exponential Function}: $y_{ij}=\eta^{[y]}_{0i}+\eta^{[y]}_{1i}\times(1-\exp(-b\times {t_{ij}}))+\epsilon^{[y]}_{ij}$}\\
\hline
\multirow{3}{*}{$\boldsymbol{\eta}^{[y]}_{i}=\begin{pmatrix}\eta^{[y]}_{0i} & \eta^{[y]}_{1i} \end{pmatrix}$} & 
\multirow{3}{*}{$\boldsymbol{\Lambda}^{[y]}_{i}=\begin{pmatrix}1 & 1-\exp(-b\times {t_{ij}}) \end{pmatrix}$} & 
$\eta^{[y]}_{0i}$: the initial status \\
& & $\eta^{[y]}_{1i}$: the change from initial status to asymptotic level \\
& & $\exp(-b\times(t_{j}-t_{j-1}))$\tnote{c}: the ratio of rate-of-change at $t_{ij}$ to that at $t_{i(j-1)}$ \\
\hline
\hline
\multicolumn{3}{l}{\textbf{Jenss-Bayley Function}: $y_{ij}=\eta^{[y]}_{0i}+\eta^{[y]}_{1i}\times t_{ij}+\eta^{[y]}_{2i}\times(\exp(c\times t_{ij})-1)+\epsilon^{[y]}_{ij}$}\\
\hline
\multirow{4}{*}{$\boldsymbol{\eta}^{[y]}_{i}=\begin{pmatrix}\eta^{[y]}_{0i} & \eta^{[y]}_{1i} & \eta^{[y]}_{2i} \end{pmatrix}$} & 
\multirow{4}{*}{$\boldsymbol{\Lambda}^{[y]}_{i}=\begin{pmatrix}1 & t_{ij} & \exp(c\times {t_{ij}}-1) \end{pmatrix}$} & 
$\eta^{[y]}_{0i}$: the initial status \\
& & $\eta^{[y]}_{1i}$: the slope of linear asymptote \\
& & $\eta^{[y]}_{2i}$: the change from initial status to the linear asymptote intercept\\
& & $\exp(c\times(t_{j}-t_{j-1}))$\tnote{c}: the ratio of acceleration at $t_{ij}$ to that at $t_{i(j-1)}$ \\
\hline
\hline
\multicolumn{3}{l}{\textbf{Bilinear Spline Function with a Fixed Knot}\tnote{d}: $y_{ij}=\begin{cases}
\eta^{[y]}_{0i}+\eta^{[y]}_{1i}\times t_{ij}+\epsilon^{[y]}_{ij}, &  t_{ij}<\gamma \\
\eta^{[y]}_{0i}+\eta^{[y]}_{1i}\times \gamma+\eta^{[y]}_{2i}\times(t_{ij}-\gamma)+\epsilon^{[y]}_{ij}, & t_{ij}\ge\gamma \\
\end{cases}$}\\
\hline
\multirow{4}{*}{$\boldsymbol{\eta}^{[y]'}_{i}=\begin{pmatrix}\eta^{[y]}_{0i}+\gamma\eta^{[y]}_{1i} & \frac{\eta^{[y]}_{1i}+\eta^{[y]}_{2i}}{2} & \frac{\eta^{[y]}_{2i}-\eta^{[y]}_{1i}}{2}\end{pmatrix}$} & 
\multirow{4}{*}{$\boldsymbol{\Lambda}^{[y]'}_{i}=\begin{pmatrix}1 & t_{ij}-\gamma & |t_{ij}-\gamma| \end{pmatrix}$} & 
$\eta^{[y]}_{0i}$: the initial status \\
& & $\eta^{[y]}_{1i}$: the slope of the first linear piece \\
& & $\eta^{[y]}_{2i}$: the slope of the second linear piece \\
& & $\gamma$\tnote{c}: the transition time from $1^{st}$ linear piece to $2^{nd}$ linear piece (i.e., knot) \\
\hline
\hline
\end{tabular}
\label{tbl:traj_summary}
\begin{tablenotes}
\small
\item[a] {Growth factors $\boldsymbol{\eta}^{[y]}_{i}$ is a $K\times 1$ matrix of growth factors, where $K$ is the number of growth factors.}\\
\item[b] {Factor loadings $\boldsymbol{\Lambda}^{[y]}_{i}$ is a $J\times K$ matrix of factor loadings, where $K$ is the number of growth factors and $J$ is the number of repeated measures.}\\
\item[c] {The coefficient $b$, $c$, and $\gamma$ can also be at the individual level and be viewed as an additional growth factor in the negative exponential function, Jenss-Bayley function, and bilinear spline function, respectively. Growth curves with such additional growth factors can be modeled through the Taylor series expansion in the structural equation modeling framework. Technical details can be found in earlier studies such as \citet{Preacher2015repara}, \citet{Liu2019BLSGM} and \citet[Chapter~11]{Grimm2016growth}.}\\
\item[d] {The bilinear spline function, also referred to as the linear-linear functional form, is widely employed to describe a longitudinal process in multiple domains, such as earlier-stage and later-stage growth in intellectual development or short-term recovery period and long-term recovery period in psychotherapy. The transition time from the first to the second stage can be viewed as a free parameter. With the reparameterization shown in the table ($\boldsymbol{\eta}^{[y]'}_{i}$ and $\boldsymbol{\Lambda}^{[y]'}_{i}$ are the reparameterized growth factors and corresponding factor loadings), we can unify the expression pre- and post-knot and fit the model in the structural equation modeling framework. More technical details can be found in \citet{Liu2019BLSGM}.} 
\end{tablenotes}
\end{threeparttable}
\end{sidewaystable*}

\begin{table*}
\centering
\begin{threeparttable}
\caption{Performance Measures for Evaluating an Estimate ($\hat{\theta}$) of Parameter ($\theta$)}
\begin{tabular}{p{4cm}p{4.5cm}p{5cm}}
\hline
\hline
\textbf{Criteria} & \textbf{Definition} & \textbf{Estimate} \\
\hline
Relative Bias & $E_{\hat{\theta}}(\hat{\theta}-\theta)/\theta$ & $\sum_{s=1}^{S}(\hat{\theta}_{s}\tnote{a}-\theta)/\theta S\tnote{b}$ \\
Empirical SE & $\sqrt{Var(\hat{\theta})}$ & $\sqrt{\sum_{s=1}^{S}(\hat{\theta}_{s}-\bar{\theta}\tnote{c})^{2}/(S-1)}$ \\
Relative RMSE & $\sqrt{E_{\hat{\theta}}(\hat{\theta}-\theta)^{2}}/\theta$ & $\sqrt{\sum_{s=1}^{S}(\hat{\theta}_{s}-\theta)^{2}/S}/\theta$ \\
Coverage Probability & $Pr(\hat{\theta}_{\text{lower}}\le\theta\le\hat{\theta}_{\text{upper}})$ & $\sum_{s=1}^{S}I(\hat{\theta}_{\text{lower},s}\le\theta\le\hat{\theta}_{\text{upper},s})\tnote{d}/S$\\
\hline
\hline
\end{tabular}
\label{tbl:metric}
\begin{tablenotes}
\small
\item[a] {$\hat{\theta}_{s}$: the estimate of $\theta$ from the $s^{th}$ replication}\\ 
\item[b] {$S$: the number of replications and set as $1,000$ in our simulation study}\\
\item[c] {$\bar{\theta}$: the mean of $\hat{\theta}_{s}$'s across replications}\\
\item[d] {$I()$: an indicator function}
\end{tablenotes}
\end{threeparttable}
\end{table*}

\begin{table*}
\centering
\resizebox{1,0\textwidth}{!}{
\begin{threeparttable}
\setlength{\tabcolsep}{4pt}
\renewcommand{\arraystretch}{0.8}
\caption{Simulation Design for TVC Decomposition Methods with Individual Measurement Occasions}
\begin{tabular}{p{4cm} p{10.5cm}}
\hline
\hline
\multicolumn{2}{c}{\textbf{Sample Size and common conditions across longitudinal processes}}\\
\hline
Sample size & $n=200, 500$ \\
\hline
\multirow{2}{*}{Study wave ($t_{j}$)} & $6$ equally-spaced: $t_{j}=0, 1, \dots, J-1\quad(J=6)$\\
& $10$ equally-spaced: $t_{j}=0, 1.00, \dots, J-1\quad(J=10)$\\
\hline
Individual time ($t_{ij}$) & $t_{ij} \sim U(t_{j}-\Delta, t_{j}+\Delta)$ ($\Delta=0.25$) \\
\hline
\hline
\multicolumn{2}{c}{\textbf{Parameters of the growth curves of the longitudinal outcome}}\\
\hline
\hline
Mean vector & $\begin{pmatrix} \mu^{[y]}_{\eta_{0}} & \mu^{[y]}_{\eta_{1}} & \mu^{[y]}_{\eta_{2}} \end{pmatrix}=\begin{pmatrix} 100 & 5 & 1.8 \end{pmatrix}$\\
\hline
\multirow{3}{*}{Variance-covariance matrix} & $\begin{pmatrix} \psi^{[y]}_{00} & \psi^{[y]}_{01} & \psi^{[y]}_{02} \\ \psi^{[y]}_{01} & \psi^{[y]}_{11} & \psi^{[y]}_{12} \\ \psi^{[y]}_{02} & \psi^{[y]}_{12} & \psi^{[y]}_{22} \end{pmatrix}=\begin{pmatrix} 25 & 1.5 & 1.5 \\
1.5 & 1.0 & 0.3 \\ 1.5 & 0.3 & 1.0 \end{pmatrix}$\\
\hline
\multirow{2}{*}{Knot location} & $\gamma=2.5$ when $J=6$ \\
& $\gamma=4.5$ when $J=10$ \\
\hline
Residual variance & $\theta^{[y]}_{\epsilon}=1$ or $2$ \\
\hline
\hline
\multicolumn{2}{c}{\textbf{Parameters of the growth curves of the time-varying covariate}}\\
\hline
\hline
\textbf{Variables} & \textbf{Conditions} \\
\hline
Mean vector & $\begin{pmatrix} \mu^{[x]}_{\eta_{0}} & \mu^{[x]}_{\eta_{1}} \end{pmatrix}=\begin{pmatrix} 10 & 5 \end{pmatrix}$\\
\hline
\multirow{2}{*}{Variance-covariance matrix} & $\begin{pmatrix} \psi^{[x]}_{00} & \psi^{[x]}_{01} \\ \psi^{[x]}_{01} & \psi^{[x]}_{11} \end{pmatrix}=\begin{pmatrix} 16 & 1.2 \\
1.2 & 1.0 \end{pmatrix}$\\
\hline
\multirow{2}{*}{Relative Rate-of-Change\tnote{a}} & $6$ waves: $\gamma_{1}=1.0$ (fixed), $\gamma_{2/3/4/5}=0.9/0.8/0.7/0.6$ \\
& $10$ waves: $\gamma_{1}=1.0$ (fixed), $\gamma_{2/3/4/5/6/7/8/9}=0.9/0.8/0.7/0.6/0.5/0.4/0.3/0.2$ \\
\hline
Residual variance & $\theta^{[x]}_{\epsilon}=1$ \\
\hline
\hline
\multicolumn{2}{c}{\textbf{Parameters related to time-invariant covariate and initial trait}}\\
\hline
\hline
\textbf{Variables} & \textbf{Conditions} \\
\hline
Correlation\tnote{b} & $\rho_{BL}=0.3$ \\
\hline
\multirow{6}{*}{Coef. to growth factors} & Time-invariant covariate explains $13\%$ variability of growth factors \\ & initial trait explains $0\%$ variability of growth factors \\
\cline{2-2}
& Time-invariant covariate explains $3\%$ variability of growth factors \\ & initial trait explains $7\%$ variability of growth factors\tnote{b} \\
\cline{2-2}
& Time-invariant covariate explains $6\%$ variability of growth factors \\ & initial trait explains $14\%$ variability of growth factors\tnote{b} \\
\hline
Time-invariant covariate & $X\sim N(0, 1^{2})$ \\
\hline
\hline
\multicolumn{2}{c}{\textbf{Parameters related temporal states}}\\
\hline
\hline
\textbf{Variables} & \textbf{Conditions} \\
\hline
Decomposition method 1 & $\kappa_{1}=0$ or $0.2$ or $0.4$ or $0.6$ \\
Decomposition method 2 & $\kappa_{2}=0$ or $0.2$ or $0.4$ or $0.6$ \\
Decomposition method 3 & $\kappa_{3}=0$ or $0.2$ or $0.4$ or $0.6$ \\
\hline
\hline
\multicolumn{2}{c}{\textbf{Other parameters}}\\
\hline
\hline
\textbf{Variables} & \textbf{Conditions} \\
\hline
Residual covariance & $\theta^{[xy]}_{\epsilon}=0.3\times\sqrt{\theta^{[x]}_{\epsilon}\times\theta^{[y]}_{\epsilon}}$ \\
\hline
\hline
\end{tabular}
\label{tbl:simu_design}
\begin{tablenotes}
\small
\item[a] {Relative rate-of-change is defined as the absolute rate-of-change over the shape factor.} \\
\item[b] {The correlation between the time-invariant covariate and the initial trait is wet as $0.3$ so that they together explain $13\%$ and $26\%$ in the second and third conditions of the coefficients to growth factors of the longitudinal outcome, respectively.}\\
\end{tablenotes}
\end{threeparttable}}
\end{table*}

\begin{table*}
\centering
\resizebox{1.1\textwidth}{!}{
\begin{threeparttable}
\small
\setlength{\tabcolsep}{4pt}
\renewcommand{\arraystretch}{0.75}
\caption{Summary of Model Fit Information For the Models}
\begin{tabular}{lrrrrr}
\hline
\hline
\textbf{Model} & \textbf{-2ll} & \textbf{AIC}  & \textbf{BIC}  & \textbf{\# of Para.} & \textbf{Math Res.}  \\
\hline
BLSGM\tnote{a} & $25332.96$ & $25355$ & $25399$ & $11$ & $35.88$ \\
BLSGM\tnote{a} w/ a TIC & $26451.70$ & $26484$ & $26548$ & $16$ & $35.88$ \\
BLSGM\tnote{a} w/ a TIC and a TVC & $37939.45$ & $38009$ & $38149$ & $35$ & $34.36$ \\
BLSGM\tnote{a} w/ a TIC and a Decomposed TVC (w/ Interval-specific Slopes) & $34097.35$ & $34167$ & $34307$ & $35$ & $33.56$ \\
BLSGM\tnote{a} w/ a TIC and a Decomposed TVC (w/ Interval-specific Changes) & $34096.30$ & $34166$ & $34306$ & $35$ & $33.52$ \\
BLSGM\tnote{a} w/ a TIC and a Decomposed TVC (w/ Change-from-baseline Values)& $33771.76$ & $33842$ & $33981$ & $35$ & $33.54$ \\
\hline
\hline
\end{tabular}
\label{tbl:info}
\begin{tablenotes}
\small
\item[a] BLSGM stands for bilinear spline growth model with an unknown fixed knot.\\
\end{tablenotes}
\end{threeparttable}}
\end{table*}

\begin{table*}
\centering
\resizebox{1.0\textwidth}{!}{
\begin{threeparttable}
\setlength{\tabcolsep}{4pt}
\caption{Estimates of BLSGM\tnote{a} with a TIC and a Decomposed TVC with Interval-specific Slopes }
\begin{tabular}{C{2cm}R{5cm}R{3cm}C{2cm}R{5cm}R{3cm}}
\hline
\hline
\textbf{Para.} & \textbf{Estimate (SE)} & \textbf{P value} & \textbf{Para.} & \textbf{Estimate (SE)} & \textbf{P value} \\
\hline
\multicolumn{3}{c}{\textbf{Parameters of TVC}} & \multicolumn{3}{c}{\textbf{Parameters of Outcome}} \\
\hline
$\mu^{[x]}_{\eta_{0}}$ & $0.0615$ ($0.0607$) & $0.3392$ & $\mu^{[y]}_{\eta_{0}}$ & $23.3945$ ($0.6212$) & $<0.0001^{\ast}$ \\
$\mu^{[x]}_{\eta_{1}}$ & $0.1677$ ($0.0061$) & $<0.0001^{\ast}$ & $\mu^{[y]}_{\eta_{1}}$ & $1.6274$ ($0.0153$) & $<0.0001^{\ast}$ \\
$\phi^{[x]}_{00}$ & $1.1655$ ($0.0931$) & $<0.0001^{\ast}$ & $\mu^{[y]}_{\eta_{2}}$ & $0.6389$ ($0.0201$) & $<0.0001^{\ast}$ \\
$\phi^{[x]}_{01}$ & $-0.0023$ ($0.0015$) & $0.1236$ & $\psi^{[y]}_{00}$ & $71.3255$ ($7.2811$) & $<0.0001^{\ast}$ \\
$\phi^{[x]}_{11}$ & $0.0004$ ($0.0001$) & $<0.0001^{\ast}$ & $\psi^{[y]}_{01}$ & $0.0647$ ($0.1493$) & $0.7722$ \\
$\gamma_{3}$ & $0.8135$ ($0.0535$) & $<0.0001^{\ast}$ & $\psi^{[y]}_{02}$ & $-1.2127$ ($0.1828$) & $<0.0001^{\ast}$ \\
$\gamma_{4}$ & $1.4073$ ($0.0616$) & $<0.0001^{\ast}$ & $\psi^{[y]}_{11}$ & $0.0501$ ($0.0058$) & $<0.0001^{\ast}$ \\
$\gamma_{5}$ & $0.5995$ ($0.0421$) & $<0.0001^{\ast}$ & $\psi^{[y]}_{12}$ & $-0.0063$ ($0.0056$) & $0.2828$ \\
$\gamma_{6}$ & $0.8381$ ($0.0446$) & $<0.0001^{\ast}$ & $\psi^{[y]}_{22}$ & $0.0600$ ($0.0100$) & $<0.0001^{\ast}$ \\
$\gamma_{7}$ & $0.3531$ ($0.0240$) & $<0.0001^{\ast}$ & $\gamma$ & $107.1501$ ($0.0541$) & $<0.0001^{\ast}$ \\
$\gamma_{8}$ & $0.3393$ ($0.0233$) & $<0.0001^{\ast}$ & $\theta^{[y]}_{\epsilon}$ & $33.5570$ ($0.9832$) & $<0.0001^{\ast}$ \\
\cline{4-6}
$\gamma_{9}$ & $0.3035$ ($0.0230$) & $<0.0001^{\ast}$ & \multicolumn{3}{c}{\textbf{Coef. of TIC}}  \\
\cline{4-6}
$\theta^{[x]}_{\epsilon}$ & $0.3442$ ($0.0095$) & $<0.0001^{\ast}$ & $\beta_{TIC0}$ & $0.1969$ ($0.4983$) & $0.7196$ \\
\cline{1-3}
\multicolumn{3}{c}{\textbf{Parameters of TIC}} & $\beta_{TIC1}$ & $0.0113$ ($0.0140$) & $0.4252$ \\
\cline{1-3}
$\mu_{x}$ & $0.0000$ ($0.0499$) & $>0.9999$ & $\beta_{TIC2}$ & $0.0102$ ($0.0192$) & $0.6057$ \\
\cline{4-6}
$\phi_{x}$ & $0.9975$ ($0.0705$) & $<0.0001^{\ast}$ & \multicolumn{3}{c}{\textbf{Coef. of TVC Trait}} \\
\cline{1-6}
\multicolumn{3}{c}{\textbf{Cov between TIC and TVC Trait}} & $\beta_{TVC0}$ & $6.0689$ ($0.4990$) & $<0.0001^{\ast}$ \\
\cline{1-3}
$\text{cov}_{\text{BL}}$ & $0.2506$ ($0.0568$) & $<0.0001^{\ast}$ & $\beta_{TVC1}$ & $0.0911$ ($0.0146$) & $<0.0001^{\ast}$ \\
\cline{1-3}
\multicolumn{3}{c}{\textbf{Residual Covariance}} & $\beta_{TVC2}$ & $-0.0203$ ($0.0184$) & $0.3472$ \\
\cline{1-6}
$\theta^{[xy]}_{\epsilon}$ & $0.6804$ ($0.0700$) & $<0.0001^{\ast}$ & \multicolumn{3}{c}{\textbf{Coef. of TVC State}} \\
\cline{4-6}
& & & $\kappa_{1}$ & $27.3675$ ($1.5892$) & $<0.0001^{\ast}$ \\
\hline
\hline
\end{tabular}
\label{tbl:est_TVCslp}
\begin{tablenotes}
\small
\item[a] BLSGM stands for bilinear spline growth model with an unknown fixed knot.\\
\item $^{\ast}$ indicates statistical significance at $0.05$ level. 
\end{tablenotes}
\end{threeparttable}}
\end{table*}

\begin{table*}
\centering
\resizebox{1.0\textwidth}{!}{
\begin{threeparttable}
\setlength{\tabcolsep}{4pt}
\caption{Estimates of BLSGM\tnote{a} with a TIC and a Decomposed TVC with Interval-specific Changes}
\begin{tabular}{C{2cm}R{5cm}R{3cm}C{2cm}R{5cm}R{3cm}}
\hline
\hline
\textbf{Para.} & \textbf{Estimate (SE)} & \textbf{P value} & \textbf{Para.} & \textbf{Estimate (SE)} & \textbf{P value} \\
\hline
\multicolumn{3}{c}{\textbf{Parameters of TVC}} & \multicolumn{3}{c}{\textbf{Parameters of Outcome}} \\
\hline
$\mu^{[x]}_{\eta_{0}}$ & $0.0582$ ($0.0609$) & $0.3392$ & $\mu^{[y]}_{\eta_{0}}$ & $23.6157$ ($0.6113$) & $<0.0001^{\ast}$ \\
$\mu^{[x]}_{\eta_{1}}$ & $0.1686$ ($0.0062$) & $<0.0001^{\ast}$ & $\mu^{[y]}_{\eta_{1}}$ & $1.6243$ ($0.0165$) & $<0.0001^{\ast}$ \\
$\phi^{[x]}_{00}$ & $1.1659$ ($0.0931$) & $<0.0001^{\ast}$ & $\mu^{[y]}_{\eta_{2}}$ & $0.6857$ ($0.0188$) & $<0.0001^{\ast}$ \\
$\phi^{[x]}_{01}$& $-0.0024$ ($0.0015$) & $0.1236$ & $\psi^{[y]}_{00}$ & $69.4127$ ($7.2692$) & $<0.0001^{\ast}$ \\
$\phi^{[x]}_{11}$ & $0.0005$ ($0.0001$) & $<0.0001^{\ast}$ & $\psi^{[y]}_{01}$ & $0.0453$ ($0.1563$) & $0.7722$ \\
$\gamma_{3}$ & $0.8172$ ($0.0547$) & $<0.0001^{\ast}$ & $\psi^{[y]}_{02}$ & $-1.0274$ ($0.1673$) & $<0.0001^{\ast}$ \\
$\gamma_{4}$  & $1.3825$ ($0.0615$) & $<0.0001^{\ast}$ & $\psi^{[y]}_{11}$ & $0.0542$ ($0.0064$) & $<0.0001^{\ast}$ \\
$\gamma_{5}$ & $0.6134$ ($0.0436$) & $<0.0001^{\ast}$ & $\psi^{[y]}_{12}$ & $-0.0058$ ($0.0054$) & $0.2828$ \\
$\gamma_{6}$ & $0.8165$ ($0.0435$) & $<0.0001^{\ast}$ & $\psi^{[y]}_{22}$ & $0.0521$ ($0.0084$) & $<0.0001^{\ast}$ \\
$\gamma_{7}$ & $0.3558$ ($0.0228$) & $<0.0001^{\ast}$ & $\gamma$ & $104.7100$ ($0.0583$) & $<0.0001^{\ast}$ \\
$\gamma_{8}$ & $0.3511$ ($0.0217$) & $<0.0001^{\ast}$ & $\theta^{[y]}_{\epsilon}$ & $33.5222$ ($0.9833$) & $<0.0001^{\ast}$ \\
\cline{4-6}
$\gamma_{9}$ & $0.2842$ ($0.0217$) & $<0.0001^{\ast}$ & \multicolumn{3}{c}{\textbf{Coef. of TIC}}  \\
\cline{4-6}
$\theta^{[x]}_{\epsilon}$ & $0.3444$ ($0.0096$) & $<0.0001^{\ast}$ & $\beta_{TIC0}$ & $0.1806$ ($0.5030$) & $0.7196$ \\
\cline{1-3}
\multicolumn{3}{c}{\textbf{Parameters of TIC}} & $\beta_{TIC1}$ & $0.0121$ ($0.0152$) & $0.4252$ \\
\cline{1-3}
$\mu_{x}$ & $0.0000$ ($0.0499$) & $>0.9999$ & $\beta_{TIC2}$ & $0.0091$ ($0.0177$) & $0.6057$ \\
\cline{4-6}
$\phi_{x}$ & $0.9975$ ($0.0705$) & $<0.0001^{\ast}$ & \multicolumn{3}{c}{\textbf{Coef. of TVC Trait}} \\
\cline{1-6}
\multicolumn{3}{c}{\textbf{Cov between TIC and TVC Trait}} & $\beta_{TVC0}$ & $6.0415$ ($0.4990$) & $<0.0001^{\ast}$ \\
\cline{1-3}
$\text{cov}_{\text{BL}}$ & $0.2509$ ($0.0569$) & $<0.0001^{\ast}$ & $\beta_{TVC1}$ & $0.0951$ ($0.0153$) & $<0.0001^{\ast}$ \\
\cline{1-3}
\multicolumn{3}{c}{\textbf{Residual Covariance}} & $\beta_{TVC2}$ & $-0.0159$ ($0.0169$) & $0.3472$ \\
\cline{1-6}
$\theta^{[xy]}_{\epsilon}$ & $0.6736$ ($0.0704$) & $<0.0001^{\ast}$ & \multicolumn{3}{c}{\textbf{Coef. of TVC State}} \\
\cline{4-6}
& & & $\kappa_{2}$ & $4.3736$ ($0.2870$) & $<0.0001^{\ast}$ \\
\hline
\hline
\end{tabular}
\label{tbl:est_TVCchg}
\begin{tablenotes}
\small
\item[a] BLSGM stands for bilinear spline growth model with an unknown fixed knot.\\
\item $^{\ast}$ indicates statistical significance at $0.05$ level. 
\end{tablenotes}
\end{threeparttable}}
\end{table*}

\end{document}